\documentclass[prb, aps, twocolumn, floats, showpacs, superscriptaddress]{revtex4-2}
\usepackage[latin3]{inputenc}
\usepackage[makeroom]{cancel}
\usepackage{graphicx}
\usepackage{amsmath}
\usepackage{physics}
\usepackage{dsfont}
\usepackage{epstopdf}
\usepackage{amsfonts}
\usepackage{amssymb}
\usepackage{xcolor}
\usepackage{graphics} 
\usepackage{dcolumn}
\usepackage{bm}
\usepackage{array}
\usepackage{appendix}
\usepackage{nicefrac}
\usepackage{cellspace} 
\usepackage[left=2cm,right=2cm,top=2cm,bottom=2cm]{geometry}
\usepackage[normalem]{ulem}
\RequirePackage[
   hyperindex,colorlinks,bookmarksnumbered,
   plainpages=true,pdfstartview=FitH]{hyperref}
\hypersetup{linkcolor=blue,urlcolor=blue,citecolor=blue}

\newcommand{\sgn}{\text{sgn}}

\begin{document}
\title{
Noise signatures of a charged Sachdev-Ye-Kitaev dot in mesoscopic transport}
\author{Andrei I. Pavlov}
\email{andrei.pavlov@kit.edu}
\affiliation{IQMT, Karlsruhe Institute of Technology, 76131 Karlsruhe, Germany}
\author{Mikhail N. Kiselev}
\affiliation{The Abdus Salam International Centre for Theoretical Physics, Strada Costiera 11, I-34151 Trieste, Italy}
\begin{abstract}
We investigate quantum noise in a mesoscopic quantum dot serving as a realization of the charged Sachdev-Ye-Kitaev (SYK) model weakly coupled to a fermionic lead via a tunnel contact. We find noise signatures under voltage and temperature biases that can serve as clear markers of the SYK physics in experiments with related setups. We develop a linear response theory that treats all types of noise on the same footing and generalizes a concept of transport coefficients for charge and heat currents, as well as relations between them, to equilibrium noise power. Within this theory, we find characteristic scaling of the \textit{noise coefficients} with temperature in all regimes that can be relevant for experimental realizations of the SYK dots, find a set of universal constants, with their values being unique to the SYK physics, that connect these coefficients, and characterize noise manifestations of the Coulomb blockade. Beyond SYK systems, these results may serve as a general framework for identification of non-Fermi-liquid signatures in mesoscopic transport and provide additional observables for experiments on thermoelectric phenomena.
\end{abstract}
\maketitle
\section{Introduction}
The Sachdev-Ye-Kitaev (SYK) model \cite{Sachdev1993, Kitaev2015, Sachdev2015, Gu2020, Tikhanovskaya2021} is a well-known toy model for non-Fermi-liquids that realizes the paradigm of quantum matter without quasiparticles \cite{Chowdhury2022}. It has gained extensive attention across various communities due to a unique combination of its features. This strongly interacting model is exactly solvable (in the limit of large number of particles $N$), and it possesses holographic duality with $(1+1)$-dimensional $AdS_2$ Jackiw-Teitelboim gravity \cite{Kitaev2018} (which holds even for $1/N$-corrections to the saddle point). Since random many-body interactions constitute the nature of the SYK model, it is widely used for studies of chaos and emergent ergodicity in quantum systems \cite{Jensen2016, Sonner2017, Altland2018, Altland2024}. As a strongly interacting system without quasiparticles, this model and its various generalizations are also utilized for getting insights into the physics behind strange metals \cite{Chowdhury2022}. This theoretical appeal has sparkled experimental efforts in realizing controllable systems governed by the SYK physics, as it would provide tools for measurement and observations of entities otherwise not available in experimental studies, e.g., Bekenstein-Hawking entropy of black holes \cite{Sachdev2015}. 

There are various proposals for experimental realization of the SYK model on a diverse range of platforms, e.g., Refs. \cite{Danshita2017, Pikulin2018, Chen2018, Wei2021, Baumgartner2024}. An idea of realizing the SYK physics in a mesoscopic quantum dot and detecting its signatures through transport observables (initially put forward in Ref. \cite{Chen2018}) was worked out theoretically in \cite{Gnezdilov2018, Can2019, Altland2019, Kruchkov2019, Pavlov2020, Cheipesh2020, Brzezinska2023, Shackleton2024}, and a recent experiment \cite{Anderson2024} demonstrated realistic graphene quantum dot (QD) parameters for fabrication and transport measurements in this setup. While average charge and heat currents through an SYK QD in a tunneling setup are well understood by now, fluctuations of these currents, i. e., \textit{quantum noise}, have not been addressed yet. It is well known that noise provides meaningful information about a quantum system and can be used in mesoscopic transport experiments for useful insights \cite{Landauer1996, Blanter2000, NazarovBook, Martin2005}. We perform a thorough investigation of noise features of the SYK dot coupled to a metallic lead through a tunnel contact. This investigated setup represents a typical tunneling spectroscopy probe, so our results can be immediately applied for non-invasive tunneling probes of the SYK dot.  Within a linear response theory for charge and heat currents, there are three types of noise, all carrying some information about the system: equilibrium thermal noise \cite{Landauer1993}, shot noise due to voltage bias \cite{Thielmann2003, Galperin2006, Kobayashi2021}, and delta-T noise due to temperature bias \cite{Lumbroso2018, Sivre2019, Larocque2020, Rech2020, Popoff2022, Zhang2022, Rebora2022, Iyer2023}. These types of noise can supply additional information about a system, that is not captured on the level of average currents.

We provide a unified description of noise, applicable beyond the SYK setup to an arbitrary system weakly tunnel coupled to a lead, giving detailed derivations for general results of Ref. \cite{Pavlov2025}. We discuss generalizations of the Wiedemann-Franz law to noise (zero-frequency noise power), and find universal relations between different types of noise and transport coefficients. Then we apply these results to a charged SYK dot and report the noise signatures that can be used as additional means to detect fingerprints of the SYK physics in the experimental setup. For instance, we show that information obtained from shot noise measurements can substitute results of thermoelectric measurements, adding versatility to experimental protocols. We note that despite there are a number of both experimental \cite{Dong2013, Wang2024} and theoretical \cite{Nikolaenko2023, Wang2024} works on shot noise in strange metals, to the best of our knowledge the noise inside quantum dots without quasiparticle description, that is addressed in this work, has not been analyzed so far. The general framework of \cite{Pavlov2025} could amend this, uniting both regimes as far as a system is probed through a tunnel junction.

This article has the following structure. In Sec. \ref{sec:Model}, we introduce the setup that we consider and discuss how the microscopic model is related to ongoing experimental efforts in realization of the SYK model in a graphene QD. We further specify realistic energy scales for such experiments and corresponding regimes of the SYK QD. We provide a derivation of charge and heat currents and zero noise power for the corresponding currents in Sec. \ref{sec:FCS}. Although we consider equilibrium transport under constant voltage or temperature bias, the formalism of this section can be expanded to out-of-equilibrium regimes. In Sec. \ref{sec:TrCoef}, we remind the linear response theory of thermoelectric transport and generalize the notion of Onsager transport coefficients \cite{Onsager1931} to the zero noise power. We explain the nature of non-trivial Lorenz ratios that are known for the SYK model in Sec. \ref{sec:LorNb} and calculate there the specific values of the universal relations between various transport and noise coefficients that characterize the SYK dot in different regimes (conformal regime with elastic tunneling, conformal regime under Coulomb blockade with inelastic tunneling, Schwarzian regime). We further analyze these universal ratios through numerical calculations in Sec.\ref{sec:SYKnoise} to provide their validity range, and give the scaling of all types of noise as functions of temperature. Sec. \ref{sec:Conc} provides the discussion of our results and their implications for ongoing experiments with the SYK dots and for problems beyond this setup.

\section{Model}
\label{sec:Model}
\begin{figure}[t!]
\center
\includegraphics[width=1.\columnwidth]{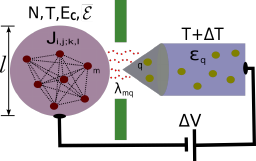} 
\caption{Considered setup: a quantum dot of size $l$ has $N$ fermions with the SYK interactions $J_{ij;kl}$ between them. This dot is at finite temperature $T$, and it possesses finite charging energy $E_C$ and spectral asymmetry $\mathcal{E}$. This system is coupled to a metallic lead with voltage bias $\Delta V$ and temperature difference $\Delta T$ through a tunnel contact with tunneling amplitudes $\lambda_{mq}$.}
\label{fig:dot}
\end{figure}

We consider a charged SYK quantum dot (SYK QD) coupled to a metallic lead via a tunnel contact, in a setup similar to \cite{Gnezdilov2018, Altland2019, Kruchkov2019, Pavlov2020, Shackleton2024}. In this setup, originally proposed in \cite{Chen2018}, a small graphene flake with highly irregular boundaries is placed into a strong magnetic field. The electrons occupying the lowest Landau level (due to a large spin splitting one may focus on the partially filled lowest Landau level for a single spin projection) become degenerate and realize the SYK interactions, while contributions from single-body terms are suppressed. This graphene flake can be weakly connected to metallic leads by tunnel contacts to probe its transport properties.  The schematic representation of the model is illustrated in Fig. \ref{fig:dot}. Here we illustrate the SYK dot coupled to a single contact, which describes a tunneling spectroscopy probe of the system under consideration. We focus on this picture since such a probe does not destroy fragile SYK physics, as elaborated below. This setup is equivalent to two identical leads coupled to the dot via tunnel contacts. One can switch between two pictures (two contacts vs one ``effective'' contact) by the means of the Glazman-Raikh rotation \cite{Glazman1988}.
The full Hamiltonian of the system reads as
\begin{align} \label{Hfull} H=H_{SYK}+E^{(0)}_C\hat{n}^{2}+H_L+H_{tun}. 
\end{align}
$H_{SYK}$ is the Hamiltonian of the charged SYK dot with $i=1..N$
electronic orbitals represented by N complex spinless
fermions $c_i$,
\begin{align} 
\label{HSYK}
H_{\rm SYK}=\frac{1}{(2N)^{3/2}}\sum_{ijkl=1}^NJ_{ij;kl}c^{\dagger}_ic^{\dagger}_jc_kc_l-\mu_D\sum_{i=1}^Nc^{\dagger}_ic_i, 
\end{align}
where $J_{ij;kl}$ are random Gaussian interaction constants (belonging to the Gaussian Unitary Ensemble, GUE) with zero mean value $\langle J_{ij;kl}\rangle = 0$ and nonzero
variance $\langle |J_{ij;kl}|^2\rangle = J^2$, these couplings are antisymmetric, $J_{ij;kl}=J^*_{kl;ij}=-J_{ji;kl}=-J_{ij;lk}$. $\mu_D$ is the chemical potential of fermions in the dot. The lead is described by the Hamiltonian 
\begin{align}
    H_L=\sum_q(\varepsilon_q-\mu_L)a^{\dagger}_qa_q,
\end{align}
where $a_q$ are fermionic operators of the lead, with momentum $q$, dispersion $\varepsilon_q$, and chemical potential $\mu_L$. Here we consider a lead with a single ballistic channel, rather than a closely related multichannel setup \cite{Can2019, Cheipesh2020}. \\
$\hat{n}=\sum_{i=1}^Nc^{\dagger}_ic_i$ is the charge of the dot, $E^{(0)}_C$ is the charging energy. This charging energy can be estimated in the following way \cite{Pavlov2020}: in a realistic quantum dot, its charging energy $E^{(0)}_C=\frac{e^2}{2C_0}$ (where $C_0$ is the dot's capacitance) scales linearly with the dot's size $l$, so $E^{(0)}_C\sim l^{-1}$. For the SYK QD, this charging energy is effectively renormalized as $E_C=E^{(0)}_C+\mathcal{K}$, where $\mathcal{K}\sim J/\left(N \ln N\right)$ for $N\gg 1$ \cite{Altland2019}. This renormalization of the charging energy also means that effects of the Coulomb blockade are always relevant in the Schwarzian regime of the SYK QD (i.e., below the $\mathcal{K}$ energy scale) \cite{Pavlov2020}.
The spectral asymmetry parameter $\mathcal{E}$ determines the asymmetry between the particle and hole excitations of the non-Fermi liquid, and is related (via the auxiliary angle $\theta$) to the total charge of the SYK dot \cite{Sachdev2015, Davison2017, Gu2020} as
\begin{align}
\label{SpAsym}    \frac{\left\langle\hat{n}\right\rangle}{N}=\frac{1}{2}-\frac{\theta}{\pi}-\frac{\sin2\theta}{4}, \,\,\, e^{2\pi \mathcal{E}}=\frac{\sin\left(\frac{\pi}{4}+\theta\right)}{\sin\left(\frac{\pi}{4}-\theta\right)}.
\end{align}

The tunneling term $H_{tun}$ reads as
\begin{align}
\label{tunneling} H_{tun}=\sum_{m=1}^N\!\sum_q\!  \frac{\lambda_{mq}}{\sqrt{2N}}a^\dag_k c_m + h.c.,
\end{align}
$\lambda_{mq}$ are random tunneling constants, we assume that they are Gaussian (GUE) with zero mean $\langle\lambda_{mq}\rangle=0$ and nonzero variance $\langle|\lambda_{mq}|^2\rangle=\lambda^2$. This assumption of random independent tunneling amplitudes for different flavors of fermions stems from random spatial structure of $c_i$ fermionic wave functions inside the dot, which allows realizing the SYK physics in such a setup in the first place \cite{Chen2018}.
We consider the weak tunneling regime, such that $\lambda$ is smaller than all relevant energy scales of the SYK dot. This smallness allows us to assume that couplings to the lead do not shift the saddle-point solution for the SYK large-$N$ action.  Furthermore, despite quadratic terms from the tunneling Hamiltonian (\ref{tunneling}) tend to destroy the conformal non-Fermi-liquid regime of the SYK dot at sufficiently low temperatures and bring the system into the Fermi-liquid state \cite{Song2017}, in the Schwarzian regime (where one accounts for renormalizations of the conformal solution by the soft mode fluctuations \cite{Bagrets2016, Maldacena2016, Kitaev2018}), the non-Fermi liquid is stable against quadratic perturbations down to arbitrary low temperatures as long as $\lambda \lesssim J/N$ \cite{Lunkin2018, Altland2019c, Altland2019, Lunkin2020, Khveshchenko2022}. Due to the same reason, small single-body perturbations $\epsilon_{ij}c^{\dagger}_ic_j$ inside the dot (that are unavoidable in realistic experiments) should not destabilize and destroy the SYK non-Fermi liquid behavior if their energies are below this threshold, so they can be neglected \cite{Altland2019}.\\
Before proceeding further, let us estimate realistic values of the parameters for possible realizations of the SYK physics in a disordered graphene quantum dot. In the original proposal \cite{Chen2018} and a related later work \cite{Brzezinska2023}, the parameters of this setup are estimated as $l\sim 100$ nm, $J~\sim 300-580$ K ($25-50$ meV).  Achievable numbers of fermions participating in Eq. (\ref{HSYK}) are expected to be around $N\sim 20-45$ (at magnetic fields $B\sim 10-40$ T; while the upper bound of this range requires too strong magnetic field for realistic devices, the lower range is easily achievable).
In a recent experiment with a disordered graphene quantum dot \cite{Anderson2024}, the number of states that might participate in the SYK interactions Eq. (\ref{HSYK}) in such a system was estimated as $N\simeq 33$ (achieved at the magnetic field $B=10$ T), the graphene island is $l\sim 100$ nm in diameter, and the measurements are performed in the temperature range of $T\sim 1.4-32$ K. It gives us an estimate for the Schwarzian energy scale $E_{Sch}\sim 3-5$ K, and suggests that this scale is relevant for experiments. In order to observe the low-temperature SYK physics, the tunneling amplitudes must be $\lambda\ll 10$ K, the renormalized charging energy is bounded from below as $E_C\gtrsim 3$ K (but it also can be much larger). According to exact diagonalization results of \cite{Bagrets2016, Lunkin2018}, the SYK two-point correlators for $N=24-32$ acquire noticeable finite-size corrections, but their structure generally reproduces the analytical predictions for both conformal and Schwarzian regimes.

\section{Full counting statistics for the charged SYK dot}
\label{sec:FCS}
The charge and heat currents emerging in the transport setup in Fig. \ref{fig:dot} of the SYK dot tunnel coupled to a lead were extensively studied in various regimes, so the manifestations of the conformal and Schwarzian regimes, their combinations with effects of the Coulomb blockade and the role of elastic and inelastic tunneling in the transport coefficients for the currents are well understood by now \cite{Altland2019, Kruchkov2019, Pavlov2020}. On the contrary, noise of these currents, which can be equally important transport signatures for experiments, have not been addressed yet, with an exception of the work \cite{Gnezdilov2018}, where the shot noise for the conformal regime ($N\rightarrow \infty$) was studied in the limit of zero temperature and without effects of the Coulomb blockade. We fill this gap by adding analysis of all kinds of noise in the corresponding transport regimes. For that, we apply a formalism of the full counting statistics (FCS) \cite{Levitov1993, Levitov1996, Nazarov2002, Bagrets2003, Nazarov2003}, allowing us to treat the charge current and noise on the same footing. This FCS approach to an SYK dot coupled to a lead was applied in, e.g., \cite{Gnezdilov2018, Cheipesh2020}. Furthermore, we want to incorporate counting of transferred heat into the same description, to get simultaneous access to statistics of transferred charge, heat and their arbitrary mixtures. In particular, we are interested in the noise of charge transfer $S_c$, of heat transfer $S_h$ and their mixture $S_m$. The generalization of the FCS approach to the heat transfer is known in literature \cite{Kindermann2004, Golubev2013, Wollfarth2014, Wollfarth2017}, as well as its applications to a general case for simultaneous charge and heat transfers \cite{Pilgram2003, Pilgram2004, Laakso2010a, Laakso2010b, Laakso2012}.\\
We consider an adiabatic steady state regime for transport, so changes of total charge and energy of the dot are equal to changes of the same quantities of the lead, $\langle \dot{Q}\rangle$ and $\langle \dot{H_L} \rangle$ \cite{Pekola2021}. Introducing the total number of electrons in the lead as $n_L=\sum_q a_q^{\dagger}a_q$, we have charge and heat currents (we put $\hbar=k_B=1$)
\begin{align}
    &I_c=-\textit{i}\left[e n_L,H\right]=\textit{i}e\sum_{l,q}\left(\frac{\lambda_{lq}}{\sqrt{2N}}a^{\dagger}_qc_l-h.c.\right),&\\
    &I_h=-\textit{i}\left[H_L,H\right]=\textit{i}\sum_{l,q}(\varepsilon_q-\mu_L)\left(\frac{\lambda_{lq}}{\sqrt{2N}}a^{\dagger}_qc_l-h.c.\right).&
\end{align}
Now we introduce two independent counting fields $\chi$ and $\xi$ that detect passage of charge $e$ and heat (energy shifted by chemical potential \cite{Benenti2017, Pekola2021}) $\varepsilon_q-\mu_L$ through the tunnel junction. The counting fields are built by $H(\chi, \xi)=H+\frac{1}{2}\chi I_c+\frac{1}{2}\xi I_E $ \cite{Gnezdilov2018, Karki2018}. 

Following \cite{Cheipesh2020, Karki2018}, we derive the tunneling action within the Keldysh formalism \cite{Kamenev2011}. The generating functional $Z[\chi,\xi]$ provides us with any moment $\mathcal{C}_{n, m}$ of charge and energy transfer.
These moments are given by
\begin{align}  \label{Moments} &\mathcal{C}_{n,m}=(-\textit{i})^{n+m}\left.\frac{d^{n}}{d\chi^n}\frac{d^{m}}{d\xi^m}\ln Z[\chi,\xi]\right\vert_{\chi=\xi=0},&\\
\label{GenFun} &Z[\chi, \xi]=\langle T_Ce^{-\textit{i}\int_C dt H(\chi, \xi)}\rangle, &
\end{align}
where $T_C$ denotes time ordering along a Keldysh contour. $\mathbb{T}$ is the measurement time, and we assume $\mathbb{T}\rightarrow \infty$.\\
The time-dependent counting fields are given by
\begin{align}
    \chi(t)=\chi \,\theta(\mathbb{T}-t), \,\,\, \xi(t)=\xi \,\theta(\mathbb{T}-t).
\end{align}
Further, we define them on the Keldysh contour as $\chi_s(t)=s\chi(t)$, $\xi_s(t)=s\xi(t)$ (note that both counting fields are defined on the same contour). Using a gauge transformation, one can put the counting fields into the tunneling term, so the full Hamiltonian $H(\chi, \xi)$ is given by Eq. (\ref{Hfull}) with the tunneling term altered as
\begin{align}
    H_{tun}(\chi,\xi)=\sum_{l=1}^N\!\sum_q\!  \frac{\lambda_{lq}}{\sqrt{2N}}e^{\textit{i}\frac{e}{2} \chi+\textit{i}\frac{\varepsilon_q-\mu_L}{2}\xi}a^\dag_q c_l + h.c.
\end{align}
Here we explicitly kept the notation for charge $e$ to highlight the nature of both counting fields, as well as their difference. For instance, unlike the charge transfer, the energy transfer is not quantized. It means that the partition function $Z[\chi, \xi]$ and its cumulants have different analytical properties with respect to the $\chi$ and $\xi$ fields \cite{Kindermann2004}. 
We derive the effective tunneling action for the system in Appendix \ref{AppA}.\\
We consider this action up to the second order in $\lambda$, which reads as
\begin{align}
    S_{eff}[\chi, \xi]=-\lambda^2\int \frac{d\varepsilon}{2\pi} \Tr \left[\left\langle\hat{\mathcal{G}}^{(h,\varphi)}(\varepsilon,\chi,\xi)\right\rangle_{h, \varphi}\hat{Q}(\varepsilon)\right].
\end{align}
Here, the matrix $\hat{Q}(\varepsilon)$ is the Fourier transformed Green's function of free fermions of the lead in the $2\times 2$ Keldysh representation. The matrix $\hat{\mathcal{G}}$ contains Keldysh Green's functions of the SYK dot, as detailed in Appendix \ref{AppA}. $\left\langle...\right\rangle_{h,\varphi}$ brackets stand for averaging over the reparametrization soft mode $h$ (Schwarzian action) and the $U(1)$ field $\varphi$ (Coulomb action). The off-diagonal elements of this matrix also contain dependence on the counting fields $\chi$ and $\xi$. This immediately provides us with all moments of $\chi$ and $\xi$ distributions in accordance with Eq. (\ref{Moments}):\\
\begin{align}
\label{cumulants}    &\mathcal{C}_{n,m}=\lambda^2\int d\varepsilon  \left[\left\langle\mathcal{G}_{+-}^{(h,\varphi)}(\varepsilon,0,0)\right\rangle_{h, \varphi}Q_{-+}(\varepsilon)+\right.&\\
\nonumber &\left.(-1)^{n+m}\left\langle\mathcal{G}_{-+}^{(h,\varphi)}(\varepsilon,0,0)\right\rangle_{h, \varphi}Q_{+-}(\varepsilon) \right]e^n(\varepsilon-\mu_L)^m.&
\end{align}
Among these moments, we consider charge current $I_c=\mathcal{C}_{1,0}$, heat current $I_h=\mathcal{C}_{0,1}$, charge noise $S_c=\mathcal{C}_{2,0}$, mixed (cross-correlated) noise $S_m=\mathcal{C}_{1,1}$, and heat noise $S_h=\mathcal{C}_{0,2}$. From now on, we put $e=1$.

\section{Transport coefficients and noise relations}
\label{sec:TrCoef}
Now, we explicitly write expressions for transport coefficients and all types of noise and bring them into the form presented in \cite{Pavlov2025}. The detailed derivations are given in Appendix \ref{AppB}. In the following we concentrate on the linear response regime since voltage and temperature biases are assumed to be small, to keep the regime of equilibrium current well justified.\\
Transport coefficients for charge and heat currents, $I_c$ and $I_h$, under applied voltage and temperature biases, $\Delta V$ and $\Delta T$, in the linear response regime are
\begin{align}
\label{TransportCoeff}    \begin{pmatrix}
        I_c \\ I_h
    \end{pmatrix}=\begin{pmatrix}
        G & G_T\\
        TG_T & G_H
    \end{pmatrix} \begin{pmatrix}
        \Delta V \\ \Delta T
    \end{pmatrix}.
\end{align}
$G$ is electric conductance, $G_T$ is thermoelectric coefficient, $G_H$ is heat conductance, $K=G_H-TG_T^2/G$ is thermal conductance \cite{Onsager1931, Costi2010}. For the considered setup, they have been analyzed previously in \cite{Pavlov2020} for all possible transport cases (elastic and inelastic transport with Coulomb blockade in conformal and Schwarzian regimes), but we explicitly provide them here for consistency, since they will be connected with various types of noise. \\
The density of states (DoS) of the dot can be expressed through the dot's transmission coefficient (imaginary part of the T-matrix, which we further refer simply as the T-matrix for brevity) in the Matsubara representation $\mathcal{T}(\tau)$ ($\tau$ is the Matsubara time) \cite{Matveev2002}. This allows writing the transport coefficients in terms of the T-matrix of the dot in a general case as
\begin{align}
\label{Gcoeff}    &G=-\frac{1}{2v_F}\int_{-\infty}^{\infty} dt \frac{1}{\cosh(\pi T t)}\mathcal{T}\left(\frac{1}{2T}+\textit{i}t\right),&\\
    &G_T=-\frac{\textit{i}\pi}{2v_F}\int_{-\infty}^{\infty} dt \frac{\sinh (\pi T t)}{\cosh^2 (\pi T t)}\mathcal{T}\left(\frac{1}{2T}+\textit{i}t\right),&\\
\label{Kcoeff}    &G_H=-\frac{\pi^2 T}{v_F}\int_{-\infty}^{\infty} dt \frac{1}{\cosh^3 (\pi T t)}\mathcal{T}\left(\frac{1}{2T}+\textit{i}t\right)-T\pi^2 G.&
\end{align}

There are three different kinds of noise: charge noise, which stems from charge transport; heat noise, due to heat transport; and mixed noise, that emerges due to correlations between charge and heat currents \cite{Sanchez2013, Battista2014, Crepieux2014}. Furthermore, each type of noise has three distinct regimes associated with it. The equilibrium thermal noise (i.e., Johnson-Nyquist noise) $S^{JN}_{c/m/h}(T)$ is present at finite temperature even in absence of voltage and temperature bias. The shot noise $S^{SN}_{c/m/h}$ is the excess noise, comparing to the equilibrium case, due to applied voltage bias. Delta-T noise $S^{\Delta T}_{c/m/h}$ is the excess noise due to applied temperature bias.\\
Note that there are different possible definitions of noise. We define it through the second moment of the full counting distribution (\ref{Moments}). Within this choice, the Fano factor for the charge transport, that is defined as a ratio between the charge shot noise and average charge current, is $F_c^{SN}=S_c^{SN}/I\rightarrow 1$ in the zero-temperature limit. This is a universal property of a tunnel contact, regardless of details of systems connected by such a contact (see Appendix \ref{AppB}), since all coherences are lost in such a tunneling process. Alternatively, noise can be defined as the symmetrized current-current correlator \cite{Blanter2000}, which is an equivalent definition for the case of the zero power noise \cite{Nazarov2009}. These symmetrized and non-symmetrized \cite{Eymeoud2016, Crepieux2021} definitions differ only by the prefactor 2. This factor of 2 difference, that holds for all formulas involving all kinds of noise, should be kept in mind when comparing results in literature, depending on which definition is used.

The Johnson-Nyquist noise is directly proportional to the transport coefficients of Eqs. (\ref{Gcoeff})-(\ref{Kcoeff}),
\begin{align}
\label{ScEq} &S_c^{JN}(T)=2TG,&\\ 
\label{SmEq} &S^{JN}_m(T)=2T^2G_T,&\\ 
\label{ShEq} &S^{JN}_h(T)=2T^2G_H.&
\end{align}
We denote the shot noise $\delta S_{c/m/h}^{SN}\equiv\left.\frac{\partial S_{c/m/h}}{\partial\Delta V}\right\vert_{\Delta T=0}$ and delta-T noise $\delta S_{c/m/h}^{\Delta T}\equiv\left.\frac{\partial S_{c/m/h}}{\partial\Delta T}\right\vert_{\Delta V=0}$. Overall, all types of noise at temperature $T$ in presence of voltage bias $\Delta V$ and thermal gradient $\Delta T$ in the linear response regime can be represented as
\begin{align}
\label{Smartix}  \renewcommand\arraystretch{1.8}  
     \begin{pmatrix}
        S_c\\ S_m\\ S_h
    \end{pmatrix}=
    \begin{pmatrix}
        S_c^{JN} & \delta S_c^{SN}&\delta S_c^{\Delta T}\\
        S_m^{JN} & \delta S_m^{SN}& \delta S_m^{\Delta T}\\
        S_h^{JN} & \delta S_h^{SN}& \delta S_h^{\Delta T}
    \end{pmatrix}
    \begin{pmatrix}
        1\\ \Delta V \\ \Delta T
    \end{pmatrix}.
\end{align}
Out of these nine elements describing different types of noise in Eq. (\ref{Smartix}), only four are independent. In addition to three types of the JN noise, fully defined by the transport coefficients of Eq. (\ref{TransportCoeff}), none of the components of the mixed noise are independent,\\
\scalebox{1.}{\parbox{1.\linewidth}{
\begin{align} \hspace{-0.15cm} \label{NoiseRec}
 \delta S_m^{SN}=T\, \delta S_c^{\Delta T}-S_c^{JN},\,   \delta S_h^{SN}=T\delta S_m^{\Delta T}-2S_m^{JN} 
\end{align} }}\\
so the shot noise and the delta-T noise of the charge and heat currents, supplemented by the transport coefficients, provide us with full information about transport in the system. \\
Transport coefficients (\ref{TransportCoeff}) can be compactly expressed through the transport integrals $\mathcal{L}_n$ \cite{Costi2010, Karki2020}, defined as
\begin{align}
    &\mathcal{L}_n=\frac{1}{4T v_F}\int_{-\infty}^{\infty}d\varepsilon \rho_{D}(\varepsilon)\frac{\varepsilon^n}{\cosh^2\left(\frac{\varepsilon}{2T}\right)} \, n=0,1,2,&
\end{align}
where $\rho_D(\varepsilon)$ is DoS of the dot.
We introduce \textit{noise integrals} $\mathcal{N}_n$ that provide concise expressions for noise:
\begin{align}
    &\mathcal{N}_n=\frac{1}{4T v_F}\int_{-\infty}^{\infty}d\varepsilon \rho_{D}(\varepsilon)\frac{\varepsilon^n\sinh\left(\frac{\varepsilon}{2T}\right)}{\cosh^3\left(\frac{\varepsilon}{2T}\right)}, \, n=0,1,2,3.&
\end{align}
With them, Eqs. (\ref{TransportCoeff}) and (\ref{Smartix}) can be written as
\begin{align} \renewcommand\arraystretch{1.2}
\label{LCoeff}    \begin{pmatrix}
        I_c \\ I_h
    \end{pmatrix}=\begin{pmatrix}
        \mathcal{L}_0 & \frac{1}{T}\mathcal{L}_1\\
        \mathcal{L}_1 & \frac{1}{T}\mathcal{L}_2
    \end{pmatrix} \begin{pmatrix}
        \Delta V \\ \Delta T
    \end{pmatrix},
\end{align}
\begin{align} \renewcommand\arraystretch{1.2}
   \begin{pmatrix}
        S_c\\ S_m\\ S_h
    \end{pmatrix}=
    \begin{pmatrix}
        2T\mathcal{L}_0 & \mathcal{N}_0 & \frac{1}{T}\mathcal{N}_1 \\
        2T\mathcal{L}_1 & \mathcal{N}_1-2T\mathcal{L}_0& \frac{1}{T}\mathcal{N}_2\\
        2T\mathcal{L}_2 & \mathcal{N}_2-4T\mathcal{L}_1& \frac{1}{T}\mathcal{N}_3
    \end{pmatrix}
    \begin{pmatrix}
        1\\ \Delta V \\ \Delta T
    \end{pmatrix}.
\end{align}

In the explicit forms derived in Appendix \ref{AppB}, the independent components of the noise are written through the T-matrix as
\begin{align}
        \delta S_c^{SN}=-\frac{\textit{i}T}{v_F}\int_{-\infty}^{\infty}dt\frac{t}{\cosh(\pi T t)}\mathcal{T}\left(\frac{1}{2T}+\textit{i}t\right),
\end{align}
\begin{align}        
    &\delta S_c^{\Delta T}=\frac{\pi T}{v_F}\int_{-\infty}^{\infty}dt \frac{\, t\sinh\left(\pi T t\right)}{\cosh^2\left(\pi T t\right)} \mathcal{T}\left(\frac{1}{2T}+\textit{i}t\right)+2G,&
\end{align}
\begin{align}
\nonumber  \delta S_h^{SN}=&-\frac{2\textit{i}\pi^2 T^3}{v_F}\int_{-\infty}^{\infty}dt \frac{t}{\cosh^3\left(\pi T t\right)}\mathcal{T}\left(\frac{1}{2T}+\textit{i}t\right)&\\
&-\pi^2T^2\,\delta S_c^{SN},&
\end{align} 
\begin{align}
   & \delta S_h^{\Delta T}=6TG_H+&\\
  \nonumber  &\frac{\pi^3T^3}{v_F}\int_{-\infty}^{\infty}dt \frac{\,t \sinh\left(\pi Tt\right)\left(5-\sinh^2\left(\pi Tt\right)\right)}{\cosh^4\left(\pi Tt\right)}\mathcal{T}\left(\frac{1}{2T}+\textit{i}t\right).&
\end{align} 

Any higher-order moments and their components due to charge and temperature bias also can be obtained straightforwardly from Eq. (\ref{cumulants}) by applying the prescription of Appendix \ref{AppB}. Note that we have not specified the T-matrix of the dot so far, so these results hold in general for a Fermi-liquid lead weakly tunnel coupled to a quantum dot, irrespective of a considered model. The universal transport properties that arise due to this generality are explored in \cite{Pavlov2025}.\\
Another observation, that follows from the derived expressions, is that all the kernels integrated with the T-matrix are either even or odd functions of the variable $t$. Therefore, if one decomposes the T-matrix $\mathcal{T}(\frac{1}{2T}+\textit{i}t)$ into even and odd components with respect to $t$, only one type of them contributes to the respective transport coefficients and noise.

Now, we can apply the general theory developed above to the specific case of the SYK dot. Note that our setup, detailed in Sec. \ref{sec:Model}, naturally requires the weak coupling regime of the tunnel contact to the dot, as otherwise the SYK physics is destroyed by such a probe. In the elastic tunneling case, the T-matrix of the dot in the weak tunneling regime is determined by the two-point correlation function of the dot. For the SYK dot considered in Sec. \ref{sec:FCS}, $\mathcal{T}_{et}(\tau)=\lambda^2\left\langle\mathcal{G}^{(h,\varphi)}\left(\tau,0,0\right)\right\rangle_{h,\varphi}$ in the Matsubara representation \cite{Pavlov2020}. The two-point Green's function factorizes with respect to the propagators accounting for charging effects and soft-mode fluctuations $\mathcal{T}_{el}(\tau)=\lambda^2G(\tau)D(\tau)$. Here, $D(\tau)$ is the two-point Coulomb correlator \cite{Efetov2003, Altland2019}
\begin{align}
\label{Coulomb}    D(\tau)=\frac{\theta_3\left(-\textit{i}E_C\tau-\textit{i}\mathcal{E}\pi,e^{-\frac{E_C}{T}}\right)}{\theta_3\left(-\textit{i}\mathcal{E}\pi, e^{-\frac{E_C}{T}}\right)}e^{-E_C|\tau|}.
\end{align}
In this expression, $\theta_3\left(\bullet,*\right)$ is the Jacobi $\theta$-function $\theta_3(z,q)=\sum_{-\infty}^{\infty}q^{n^2}e^{\textit{i}zn}$. In the high-temperature regime, where the charging energy is negligible, $D(\tau)\underset{T\gg E_C}{\rightarrow}1$. In the opposite limit of the dominant charging energy at low temperature $T\ll E_C$, we have the leading contribution $D(\tau)\underset{T\ll E_C}{\simeq}e^{-E_C|\tau|}$, with exponentially small corrections that depend on spectral asymmetry $\mathcal{E}$. In this low-temperature regime, inelastic tunneling processes become dominant for diagonal transport coefficients, as was first shown in \cite{Altland2019}. Although the leading inelastic contribution is $\sim \lambda^4$, and is negligible at high temperatures due to parametric smallness of the tunneling $\lambda$, it inevitably wins against exponentially suppressed elastic terms, that $\sim \lambda^2 e^{-\frac{E_C}{2T}}$, at sufficiently low temperatures. Moreover, due to the effective enhancement of the charging energy in the Schwarzian regime \cite{Altland2019}, the inelastic tunneling processes are always important below the Schwarzian energy scale, as $E_{Sch}< E_C$. The inelastic $T$-matrix is given by four-point correlators $\mathcal{T}_{in}(\tau)=\lambda^4\frac{T}{\sin\left(\pi T\tau\right)}F_{SYK}(\tau)F_C(\tau)$ [$F_C(\tau)$ is the inelastic four-point Coulomb correlator, $F_{SYK}(\tau)$ is the four-point SYK correlator].

To proceed further, we need two- and four-point correlation function of the SYK dot. In the conformal regime \cite{Sachdev2015}, we have the two-point correlator
\begin{align}
\label{Gc} \scalebox{1}{
    $G^c(\tau)=-C_{\theta}\sgn(\tau)\sin\left(\frac{\pi}{4}-\sgn(\tau)\theta \right)\left(\frac{TJ}{\sin\left(\pi T|\tau|\right)}\right)^{1/2},$}
\end{align}
$C_{\theta}=\left[(8/\pi)\cos(2\theta)\right]^{-1/4}$, and angle $\theta$ is defined in Eq. (\ref{SpAsym}). The four-point correlator (in $1/N$-leading order) \cite{Altland2019} is
\begin{align}
    F^c_{SYK}(\tau)=G^c(\tau)G^c(-\tau).
\end{align}
Crossing into the Schwarzian regime, we have 
\begin{align} \label{GSch2}
    G^{Sch}(\tau)=-\sgn(\tau)\frac{\beta^{3/2}}{\left(4\pi\right)^{1/4}}\frac{\Gamma^4\left(\frac{1}{4}\right)}{\pi}\frac{me^{-\frac{\pi^2}{\beta m}}}{|\tau|^{3/2}\left(\beta-|\tau|\right)^{3/2}},
\end{align}
\begin{align} \label{FSch4}
    F_{SYK}^{Sch}(\tau)=\frac{\beta^{3/2}\pi m^{1/2}e^{-\frac{\pi^2}{\beta m}}}{2^{3/2}|\tau|^{3/2}\left(\beta-|\tau|\right)^{3/2}},
\end{align}
with $\beta=1/T$, $m=\frac{N\ln N}{64J}\sqrt{\frac{\cos(2\theta)}{2\pi}}$, as elaborated in \cite{Bagrets2016, Mertens2017}. These expressions can be immediately used to evaluate the T-matrix in all relevant regimes and calculate the transport and noise coefficients.

\section{Lorenz ratios for the SYK dot}
\label{sec:LorNb}
The conductance and thermal conductance are related through the Wiedemann-Franz (WF) law \cite{Benenti2017} and its generalization beyond Fermi liquids \cite{vanDalum2020, Kiselev2023}, $\frac{K}{T G}=L_0 R_L$. $L_0=\frac{\pi^2}{3}$ is the Lorenz number, $R_L$ is the Lorenz ratio that is used to account for deviations from the WF law in non-Fermi liquids. It sets a relation between thermal and charge conductance in a general case. This empiric law is approximate, it becomes exact when thermoelectric effects, contributing to the thermal conductance and violating this relation, are negligible \cite{Benenti2017}. The universality of the WF law in this exact regime can be written as $\frac{\mathcal{L}_2}{T^2\mathcal{L}_0}=L_0R_L$. Expressing the transport coefficients through the Johnson-Nyquist noise Eqs. (\ref{ScEq}-\ref{ShEq}), we obtain
\begin{align}
    \frac{S_h^{JN}}{T^2 S_c^{JN}}=L_0R_L.
\end{align}

The Lorenz ratio for the SYK model differs from the one given by the Fermi liquid theory (which is $R_{L,FL}=1$). As found in \cite{Davison2017, Pavlov2020}, it takes values
\begin{align} \label{LorenzR}
    R_{L,SYK}=\begin{cases}
        \frac{3}{5}, \,\,\ E_C\ll T \ll J,\\
        \frac{3}{2}, \,\,\,  E_{Sch}\ll T\ll E_C,\\
         1.55,\,\,\, T\ll E_{Sch}.
    \end{cases}.
\end{align}
Therefore, even at equilibrium without any applied bias, the ratio of two types of the Johnson-Nyquist noise can identify a signature of the SYK physics within the quantum dot.

Now, we clarify the origin of the nontrivial Lorenz ratios of the SYK dot coupled to the free fermionic lead in different regimes, $R_{L, SYK}$, given by Eq. (\ref{LorenzR}). A general theory for universal ratios between the transport and noise coefficients is presented in \cite{Pavlov2025}.

\begin{table}[!t]
    \centering
    \resizebox{0.95\columnwidth}{!}{%
    \begin{tabular}{Sc| Sc | Sc | Sc | Sc | Sc}
   \,\,\,  & $\frac{\mathcal{L}_2}{T^2 \mathcal{L}_0}$ & $\frac{\mathcal{N}_1}{T\mathcal{L}_0}$ & $\frac{\mathcal{N}_3}{T^3\mathcal{L}_0}$ & $\frac{T \mathcal{N}_0}{\mathcal{L}_1}$ & $\frac{ \mathcal{N}_2}{T\mathcal{L}_1}$ \\ \hline
     $L_i$ ($i=0,..,4$) & $\frac{\pi^2}{3}$ & 1 & $\pi^2$ & $\frac{1}{3}$ & $\frac{\pi^2+12}{9}$ \\ \hline
     $R_{i, SYK}$ (con, el) & $\frac{3}{5}$ & $\frac{2}{3}$ & $\frac{10}{21}$ & $ 1.16$ & $ 0.64$ \\ \hline
     $R_{i,SYK}$ (con, in) & $\frac{3}{2}$ & $\frac{4}{3}$ & $\frac{28}{15}$ & {\color{gray}$0.85$} & {\color{gray}$1.40$}  \\ \hline
     $R_{i, SYK}$ (Sch, el) & {\color{gray}$ 1.06$} & {\color{gray}$ 1.03$} & {\color{gray}$ 1.13$} & {\color{gray}$1.01$} & {\color{gray}$ 0.99$}  \\ \hline
     $R_{i, SYK}$ (Sch, in) & $ 1.55$ & $ 1.35$ & $ 1.98$ & {\color{gray}$ 0.84$}  & {\color{gray}$1.42$}
    \end{tabular}}
    \caption{Extended Lorenz numbers $L_i$ ($i=0,..,4$) and Lorenz ratios for the SYK dot saturated in various regimes: conformal elastic (con, el); conformal inelastic (con, in); Schwarzian elastic (Sch, el); Schwarzian inelastic (Sch, in). Numbers in gray color denote unphysical regimes for the setup. } 
    \label{tab:ratios}
\end{table}

In addition to having nontrivial Lorenz ratios within the conformal regime, the SYK dot can depart from the ``magic" Lorenz ratios, a pattern of simple fractions relating heat and charge conductance that typically persists in quantum dot setups even beyond Fermi liquids (e.g., in Kondo and quantum Hall problems \cite{Kiselev2020, Karki2020b, Kiselev2023, Stabler2023}). For the SYK dot, this pattern is followed within the conformal regime and is broken in the Schwarzian regime, as is evident from Eq. (\ref{LorenzR}). It is not surprising, since this pattern arises for systems with conformal symmetry. In these cases, the two-point correlator as a function of the imaginary time $\tau$ scales as a power law in the zero-temperature regime $G(\tau)\sim \tau^{-\alpha}$, and the finite-temperature behavior can be easily obtained by applying a conformal map, that results into the $\sim\sin^{-\alpha}\left(\pi T \tau\right)$-like behavior \cite{Kitaev2015, Sachdev2015, Tikhanovskaya2021}.
The universality of the transport coefficient ratios emerges whenever the T-matrix can be approximated as a function of a single parameter. At sufficiently high temperatures ($T\gg E_C$) and $\mathcal{E}=0$, we have
\begin{align} \label{UnivRatio}
    \frac{\mathcal{L}_2}{T^2 \mathcal{L}_0}=\pi^2\left(\frac{2\int_{-\infty}^{\infty}dx \cosh^{-3-\alpha}x}{\int_{-\infty}^{\infty}dx \cosh^{-1-\alpha}x}  -1\right)=\frac{\pi^2}{3}\frac{3\alpha}{2+\alpha},
\end{align}
\begin{figure}[t!]
\vspace{-0.75cm}
\center
\hspace*{-0.4cm}
\includegraphics[width=1.15\linewidth]{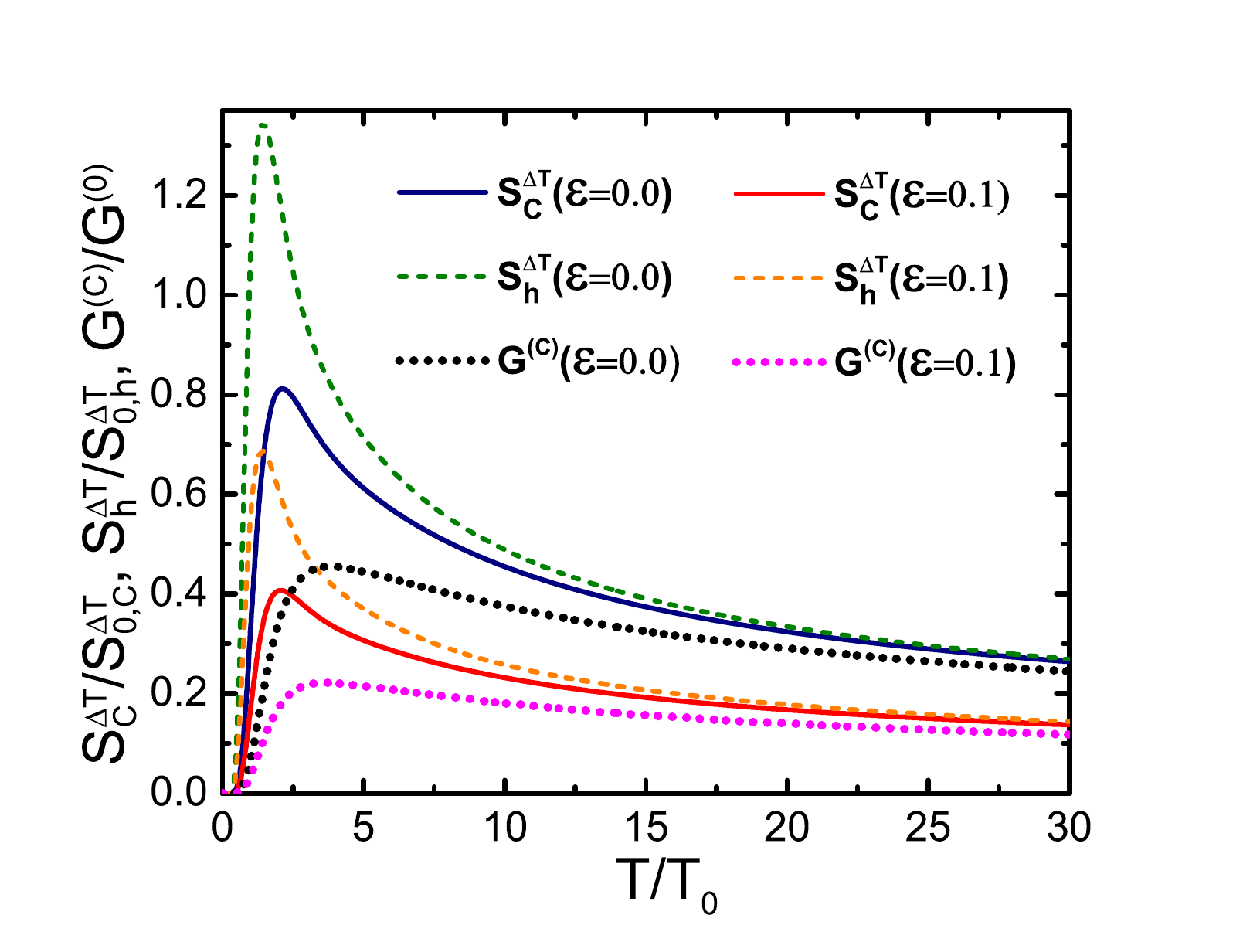} 
\vspace{-0.5cm}
\caption{Charge delta-T noise $S^{\Delta T}_c(T,\mathcal{E})$ vs heat delta-T noise $S^{\Delta T}_h(T,\mathcal{E})$ vs charge conductance $G(T,\mathcal{E})$ in the conformal regime as functions of temperature. $T_0$ is the Schwarzian energy scale, $T_0\simeq J/(N\log N)=J/100$, $E_C/T_0=10$, $N=30$. Solid blue line: $S_{c}^{\Delta T}(T,\mathcal{E}=0.0)$; solid red line: $S_{c}^{\Delta T}(T,\mathcal{E}=0.1)$; dashed green line: $S_{h}^{\Delta T}(T,\mathcal{E}=0.0)$; dashed orange line: $S_{h}^{\Delta T}(T,\mathcal{E}=0.1)$; dotted black line: $G(\mathcal{E}=0.0)$; dotted magenta line: $G(\mathcal{E}=0.1)$. Charge conductance lines are multiplied by a factor $0.67$, heat delta-T noise lines are multiplied by a factor $0.14$ (see explanation in the text). All entities with $\mathcal{E}=0.1$ are additionally multiplied by a factor $0.5$ for better visual separation. All lines are in units of $S^{\Delta T}_{0,c}=\lambda^2e^2/v_F$, $S^{\Delta T}_{0,h}=\lambda^2T^2/v_F$,$G_T^{(0)}=\lambda^2e^2/v_F$, $T_0=J/100$.}
\label{fig:DTCNc}
\end{figure}
where $x=\pi Tt$ and $\alpha=\frac{1}{2}$ (see Eq. (\ref{Gc})), and all other multipliers and constants cancel each other in the nominator and denominator. This immediately gives us $\frac{\mathcal{L}_2}{T^2 \mathcal{L}_0}=\frac{\pi^2}{5}=\frac{3}{5}L_0$. Deep within the Coulomb blockade regime, the T-matrix restores the single parametric scaling (elastic tunnelings are exponentially suppressed, so one can consider only inelastic tunneling processes). After cancellation of all prefactors, the ratio is again given by Eq. (\ref{UnivRatio}), but with $\alpha=2$, resulting in $\frac{\mathcal{L}_2}{T^2 \mathcal{L}_0}=\frac{\pi^2}{2}=\frac{3}{2}L_0$. This pattern reminds the ``magic" Lorenz ratios emerging in Kondo problems. The reason for that is the fact that in all these cases the effective T-matrix is proportional to $\mathcal{T}\sim \cosh^{-\alpha} x$ ($\alpha>1$).
For instance, the Lorenz ratio $R_L=\frac{3}{2}$ was reported \cite{Karki2020b, Kiselev2023} for two coupled $N-$ and $M-$channel Kondo simulators with $N=2$, $M\rightarrow\infty$ (interestingly, even charge conductance and heat conductance have the same scaling $\sim T$ and $\sim T^2$ correspondingly, as for the SYK dot in the inelastic conformal regime). Nevertheless, the physics behind these numbers is different. For Kondo problems, the power $\alpha$ in $\mathcal{T}\sim \cosh^{-{\alpha}}x$ is determined by DoS of the dot, which is governed by a number of channels in the system connected to the contact, and is ultimately tied to the Anderson orthogonality catastrophe \cite{Anderson1967, Mahan1967, Furusaki1995}. In contrast, for the SYK dot $\mathcal{T}\sim\cosh^{-\frac{1}{2}}x$ is set by the conformal saddle point of the model and is further renormalized by inelastic processes. This similarity ends in the Schwarzian regime, where $\mathcal{T}\sim (\frac{\pi^2}{4}+x^2)^{-\frac{3}{2}}$, departing from $\cosh$-like scaling and leading to $R_L\simeq 1.55$ (taking only elastic processes into account, this value would be $R_L\simeq 1.06$ (see Eq. (\ref{FSch4})), with the T-matrix contribution stemming from Eq. (\ref{GSch2})). 

\begin{figure}[t!]
\vspace{-0.75cm}
\center
\hspace*{-0.4cm}
\includegraphics[width=1.15\linewidth]{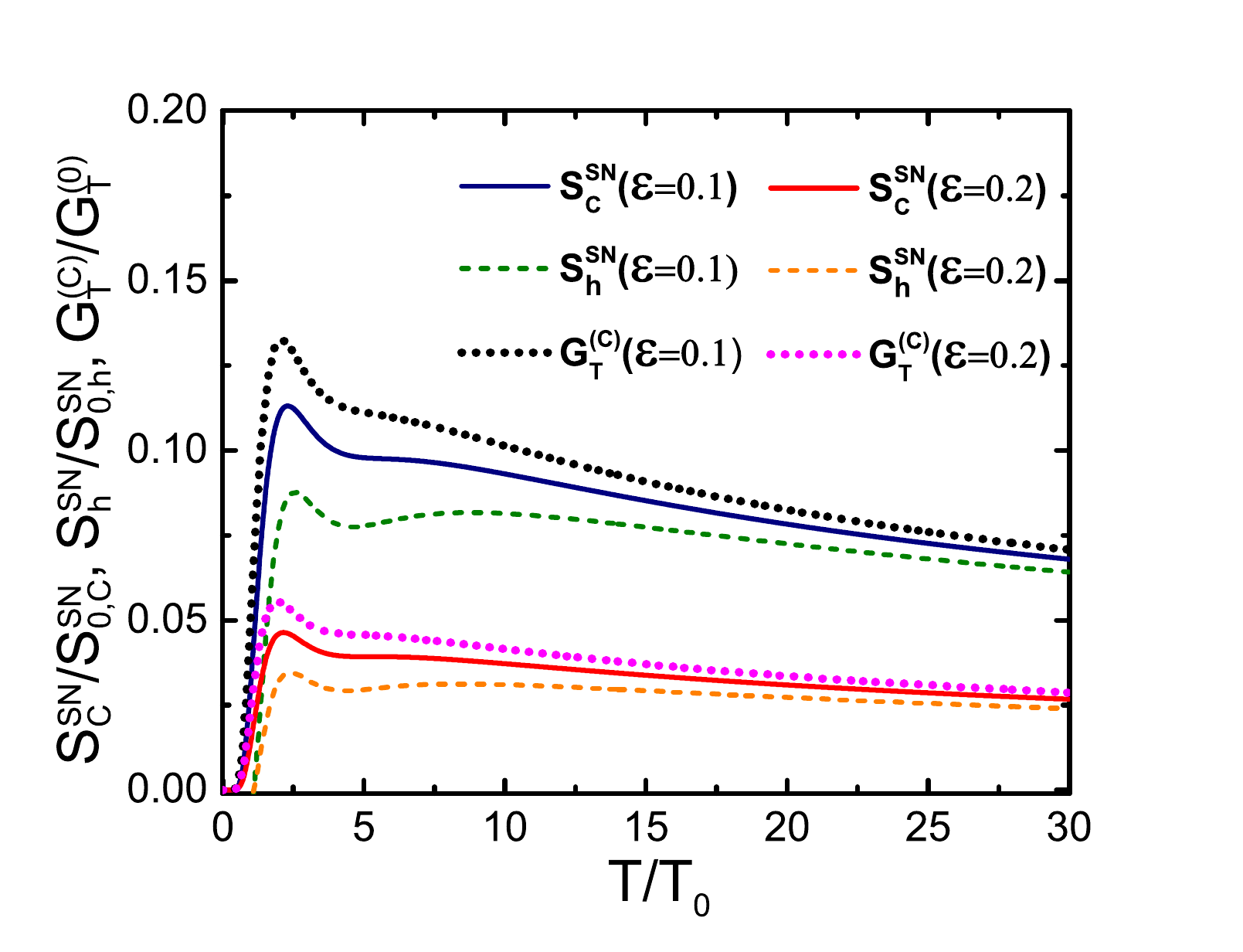} 
\vspace{-0.5cm}
\caption{Charge shot noise $S^{SN}_c(T,\mathcal{E})$ vs heat shot noise $S_h^{SN}(T,\mathcal{E})$ vs thermoelectric coefficient $G_T(T,\mathcal{E})$ in the conformal regime as functions of temperature. $T_0$ is the Schwarzian energy scale, $T_0\simeq J/(N\log N)=J/100$, $E_C/T_0=10$, $N=30$. Solid blue line: $S_{c}^{SN}(T,\mathcal{E}=0.1)$; solid red line: $S_{c}^{SN}(T,\mathcal{E}=0.2)$; dashed green line: $S_{h}^{SN}(T,\mathcal{E}=0.1)$; dashed orange line: $S_{h}^{SN}(T,\mathcal{E}=0.2)$; dotted black line: $G_T(\mathcal{E}=0.1)$; dotted magenta line: $G_T(\mathcal{E}=0.2)$. Charge shot noise lines are multiplied by a factor $2.59$, heat shot noise lines are multiplied by a factor $-0.41$ (see explanation in the text); all entities with $\mathcal{E}=0.2$ are additionally multiplied by a factor $0.5$ for better visual separation. All lines are in units of $S^{SN}_{0,c}=\lambda^2e^3/v_F$, $S^{SN}_{0,h}=\lambda^2e T^2/v_F$, $G_T^{(0)}=\lambda^2e/v_F$, $T_0=J/100$.}
\label{fig:SNGTc}
\end{figure}

\begin{figure}[t!]
\vspace{-0.75cm}
\center
\hspace*{-0.15cm}
\includegraphics[width=1.15\linewidth]{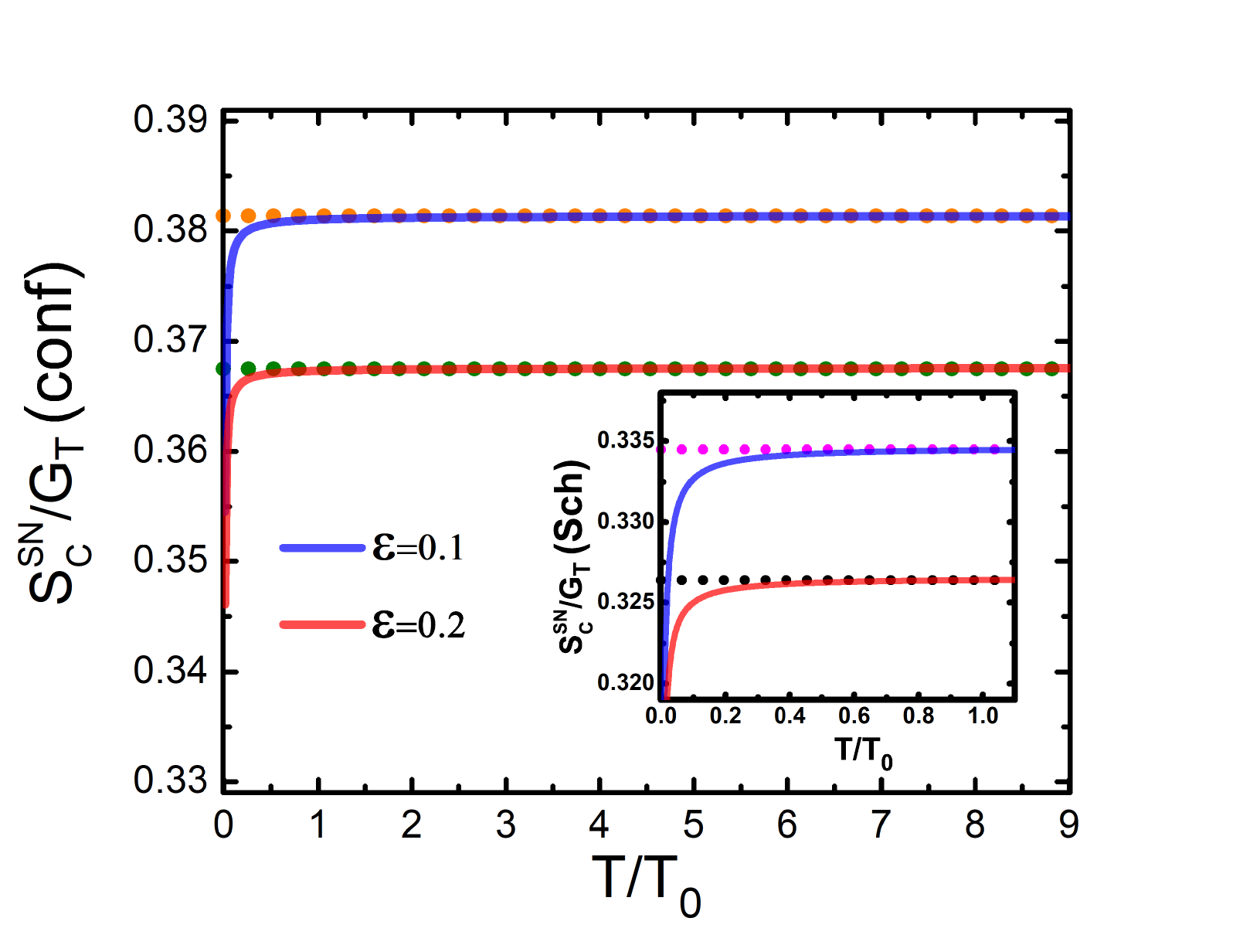} 
\vspace{-0.5cm}
\caption{Ratio between the charge shot noise $S^{SN}_c$ and the thermoelectric coefficient $G_T$ when charging energy $E_C$ is negligible. $E_C/T_0=0.01$, conformal regime ($N\rightarrow \infty$). Blue solid line - $\mathcal{E}=0.1$, red solid line - $\mathcal{E}=0.2$. Dashed lines are corresponding asymptotic values: orange dashed line is $0.3814$, green dashed line is $0.3672$. Inset: The same ratios in the Schwarzian regime (without charging energy renormalization), $N=30$, dashed magenta line is $0.3345$, dashed black line is $0.3264$.}
\label{fig:ratioSNc}
\end{figure}
Following the same approach, we calculate the ratios between charge delta-T noise, and charge conductance $\frac{\delta S_c^{\Delta T}}{G}=\frac{\mathcal{N}_1}{T\mathcal{L}_0}=L_1R_C^{\Delta T}$, heat delta-T noise and charge conductance $\frac{\delta S_h^{\Delta T}}{T^2G}=\frac{\mathcal{N}_3}{T^3\mathcal{L}_0}=L_2R_H^{\Delta T}$. Here, we introduced $L_1$, $L_2$ as the extended Lorenz numbers, setting the corresponding ratios for Fermi liquids. The coefficients $R_C^{\Delta T}$, $R_H^{\Delta T}$ account for deviations from the Fermi-liquid values, similar to the Lorenz ratio $R_L$. These values are given in Table \ref{tab:ratios}. For the antisymmetric coefficients \cite{Pavlov2025}, one can obtain the similar relations in the limit of the vanishing charging energy $E_C/T \rightarrow 0$ and spectral asymmetry $\mathcal{E}\rightarrow 0$. Although the antisymmetric coefficients do vanish in this regime, their asymptotic ratios remain finite universal values. The even components of the T-matrix do not contribute to them. The leading odd component of the T-matrix that contributes to such ratios is $\mathcal{T}_{odd}\sim \frac{\textit{i}x}{\cosh^{\alpha}\left(x\right)}$ [in the Schwarzian regime, it takes form $\mathcal{T}_{odd}\sim \textit{i}x (\frac{\pi^2}{4}+x^2)^{-\frac{3}{2}} $], the validity of this approximation will be checked in the next section. It provides the ratios $\frac{\delta S_c^{SN}}{G_T}=\frac{T\mathcal{N}_0}{\mathcal{L}_1}=L_3R_C^{V}$ and $\frac{\delta S_h^{SN}}{T^2G_T}=\frac{\mathcal{N}_2-4T\mathcal{L}_1}{T\mathcal{L}_1}=L_4R_H^{V}-4$. The numerical values of all these extensions of the Lorenz numbers and Lorenz ratios are given in Table \ref{tab:ratios}. Simple fractions listed there are exact expressions, while decimals are rounded after two digits. In this table, we list all five ratios in all possible regimes (conformal elastic, conformal inelastic, Schwarzian elastic, Schwarzian inelastic), but it is worth pointing out that only some of these regimes make physical sense. Indeed, as discussed in Sec. \ref{sec:Model}, the purely elastic Schwarzian regime is not realized due to effective renormalization of the charging energy. Furthermore, the ratios between the antisymmetric coefficients require vanishingly small charging energy. On the other hand, the purely inelastic tunneling regime can be realized only if the charging energy is sufficiently large to suppress elastic processes, so this assumption is not justified anymore. To distinguish between realistic and unphysical regimes (calculated through a formal definition), we denote the latter by the gray color in the Table \ref{tab:ratios}. Nevertheless, we note in passing that thought the elastic Schwarzian tunneling regime is not realized in the charged fermion setup, there are proposals for Majorana realization of the SYK model (e. g.,  \cite{Pikulin2018}). The WF law can be generalized to Majorana systems  (see \cite{Gnezdilov2016}), so the relations obtained for the elastic Schwarzian regime might become relevant for such types of setups.

\begin{figure}[t!]
\vspace{-0.75cm}
\center
\hspace*{-0.5cm}
\includegraphics[width=1.15\linewidth]{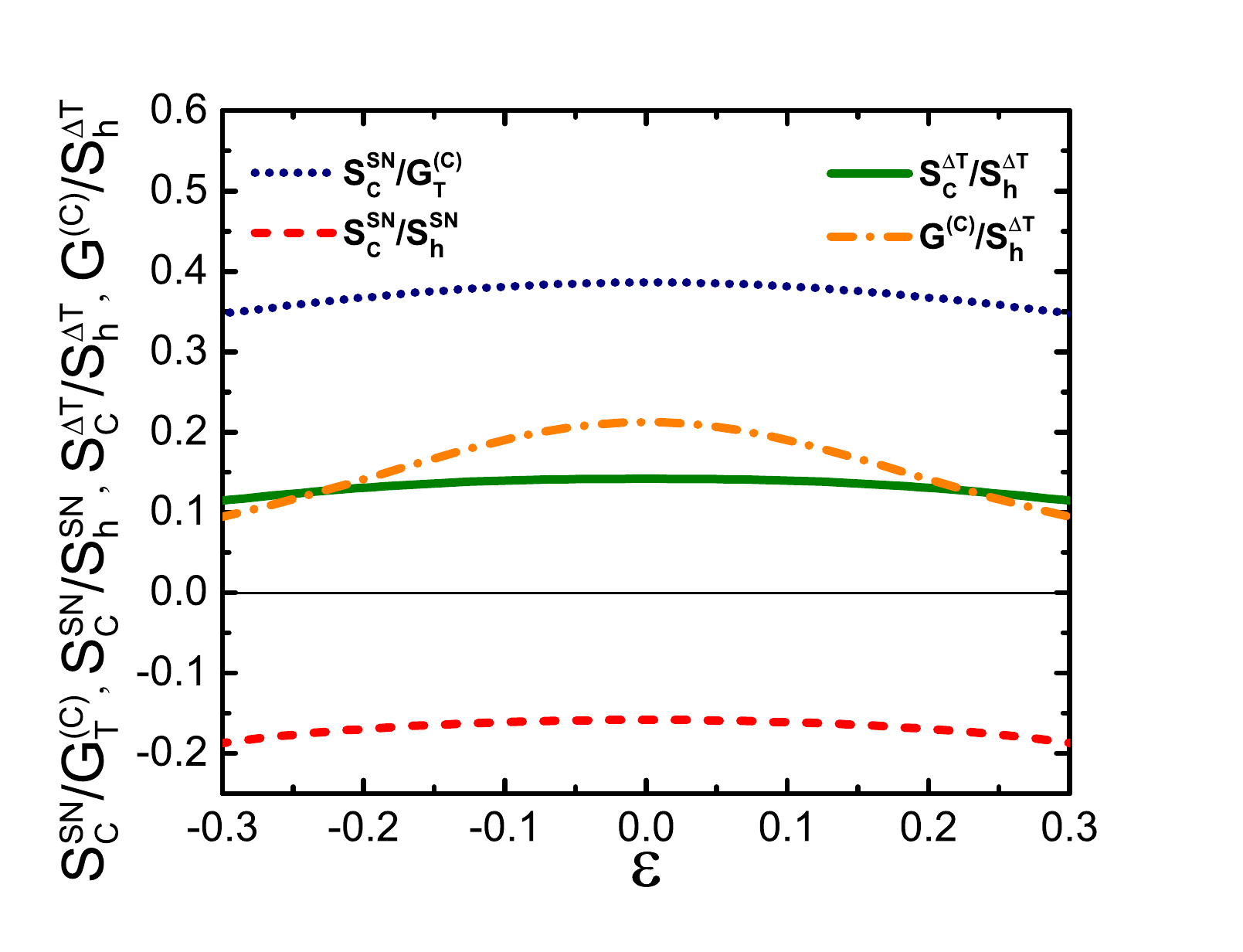} 
\vspace{-0.5cm}
\caption{Asymptotic ratios between noise and transport coefficients as functions of the spectral asymmetry $\mathcal{E}$ when charging energy $E_C$ is negligible. $T/E_C=300$, conformal regime. Blue dotted line: ratio between charge shot noise and thermoelectric coefficient, $S_{c}^{SN}/G_T$; red dashed line: ratio between charge shot noise and heat shot noise, $S_{c}^{SN}/S_{h}^{SN}$; green solid line: ratio between charge delta-T noise and heat delta-T noise, $S_c^{\Delta T}/S_h^{\Delta T}$; orange dot-dashed line: ratio between charge conductance and heat delta-T noise, $G/S_h^{\Delta T}$. All coefficients are in corresponding dimensionless units.}
\label{fig:ratiosSG_all}
\end{figure}

\section{Noise signatures of the SYK dot}
\label{sec:SYKnoise}
In this section, we focus on noise signatures for the SYK dot in various regimes and numerically check validity of the universal ratios of Sec. \ref{sec:LorNb}.

The transport coefficients for the given setup in all possible regimes were considered in \cite{Pavlov2020}. We start by summarizing these findings below. The charge conductance scales as a function of temperature \citep{Altland2019} as
\begin{align} \label{Gscaling}
    G(T)\sim\begin{cases}
        T^{-\frac{1}{2}}, \,\,\ E_C\ll T \ll J,\\
        T, \,\,\,  E_{Sch}\ll T\ll E_C,\\
        T^{\frac{3}{2}},\,\,\, T\ll E_{Sch}.
    \end{cases}
\end{align}

The thermopower (i.e.,  Seebeck coefficient) $\mathcal{S}=\frac{G_T}{G}$ is defined by the ratio between the thermoelectric coefficient and conductance, and is proportional to the spectral asymmetry $\sim \mathcal{E}$ (at small asymmetry). For the SYK dot, it scales with temperature as
\begin{align}
\label{Seds}    \mathcal{S}(T)\sim\begin{cases}
        \frac{4\pi}{3}\mathcal{E}+\mathcal{E}\mathcal{O}(1)T^{-\frac{1}{2}}, \,\,\ E_C\ll T \ll J,\\
        \mathcal{E}T^{-3}e^{-\frac{E_C}{2T}}, \,\,\,  E_{Sch}\ll T\ll E_C,\\
        \mathcal{E}T^{-\frac{7}{2}}e^{-\frac{E_C}{2T}},\,\,\, T\ll E_{Sch}.
    \end{cases}
\end{align}
The heat conductance $G_H$ behaves as
\begin{align}
    G_H(T)\sim\begin{cases}
        T^{\frac{1}{2}}, \,\,\ E_C\ll T \ll J,\\
        T^2, \,\,\,  E_{Sch}\ll T\ll E_C,\\
        T^{\frac{5}{2}},\,\,\, T\ll E_{Sch}.
    \end{cases}
\end{align}

\begin{figure}[t!]
\vspace{-0.75cm}
\center
\hspace*{-1.15cm}
\includegraphics[width=1.3\linewidth]{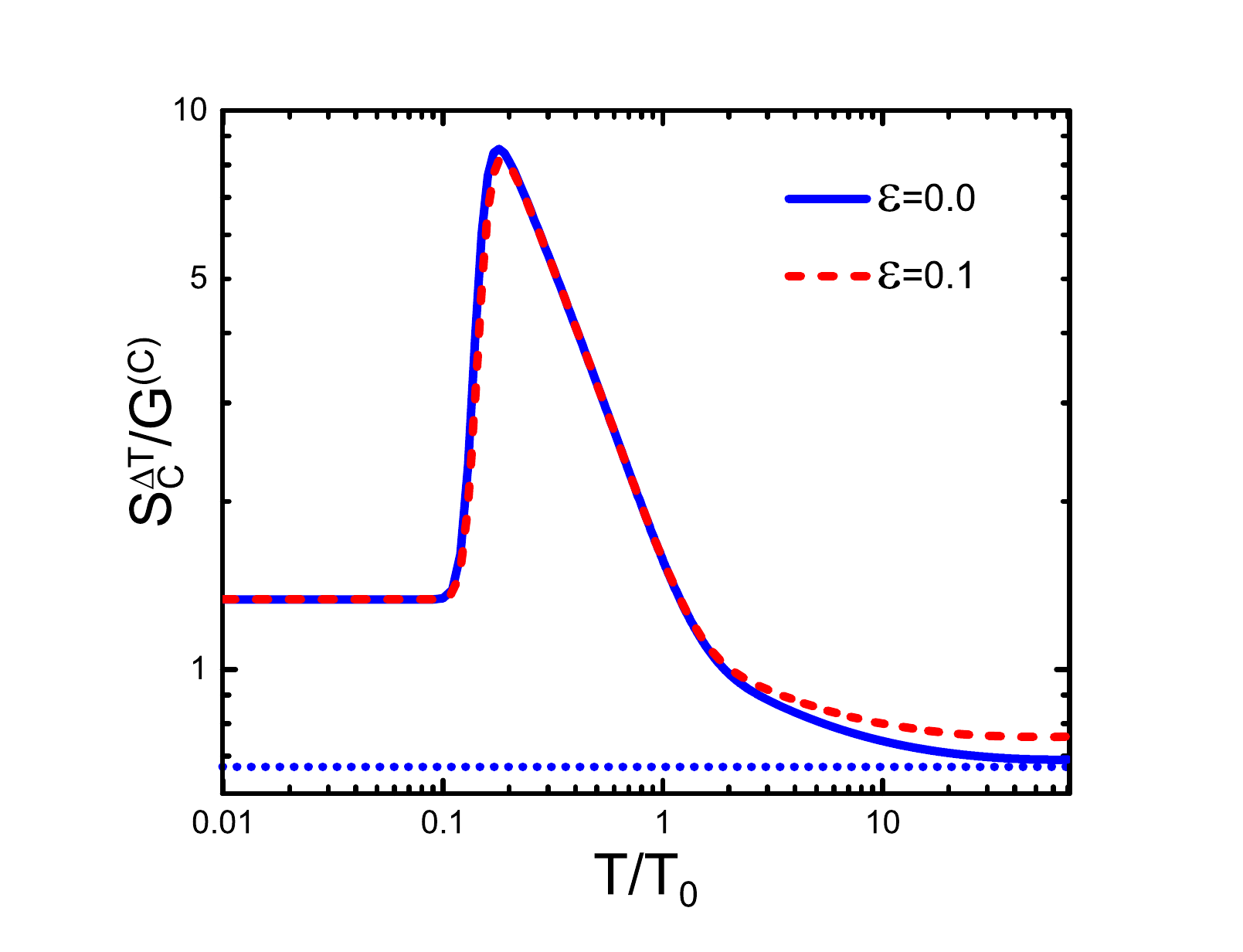} 
\vspace{-1cm}
\caption{\textit{Log-log} plot of the ratio between charge delta-T noise and charge conductance in the conformal limit ($N\rightarrow \infty$) for both elastic and inelastic processes. $\lambda/T_0=0.01$, $E_C/T_0=5$. Solid blue line: $\mathcal{E}=0.0$; dashed red line: $\mathcal{E}=0.1$. Blue dotted line corresponds to saturation value  $0.67$ of this ratio for elastic processes for $\mathcal{E}=0$ at large temperatures.}
\label{fig:ratio_NEC}
\end{figure}

Now, let us move forward and analyze four independent \textit{noise coefficients} in all regimes. 
Using the two- and four-point correlators (\ref{Coulomb}-\ref{FSch4}), one can immediately obtain explicit expressions for all types of noise and determine their scalings. Akin to transport coefficients, one half of the noise coefficients depend only on even powers of spectral asymmetry $\mathcal{E}$, which we refer as \textit{symmetric coefficients} ($\mathcal{L}_0$, $\mathcal{L}_2$, $\mathcal{N}_1$, $\mathcal{N}_3$), and the other half depend only on odd powers of spectral asymmetry $\mathcal{E}$, which we refer as \textit{antisymmetric coefficients} ($\mathcal{L}_1$, $\mathcal{N}_0$, $\mathcal{N}_2$). We plot all symmetric coefficients (conductance, charge delta-T noise and heat delta-T noise; a similar plot for the heat conductance is given in Ref. \cite{Pavlov2020}) for direct tunnelings within the conformal regime in Fig. \ref{fig:DTCNc}, and all antisymmetric coefficients (thermoelectric coefficient, charge shot noise, and heat shot noise) in Fig. \ref{fig:SNGTc}. Note that up to numerical prefactors all the coefficients (within symmetric and antisymmetric groups) have the same behavior at high temperatures (modulo additional $T^2$ for the heat noise in comparison to the charge noise), where effects of the Coulomb blockade can be neglected, while their relations are non-universal at $T\lesssim E_C$. This proportionality of the coefficients within the symmetric and antisymmetric groups, obtained here numerically, is a manifestation of the universal relations reported in \cite{Pavlov2025} and discussed in details in Sec. \ref{sec:LorNb}. The numerical prefactors, that asymptotically collapse the lines into each other (up to small nonuniversal corrections stemming from finite $\mathcal{E}$ and $E_C/T$), are taken from the analytical predictions for the conformal elastic regime. Indeed, $\frac{\delta S_c^{\Delta T}}{G}=L_1R_C^{\Delta T}\simeq 0.67$, $\frac{T^2\,\delta S_c^{\Delta T}}{\delta S_h^{\Delta T}}=\frac{L_1R_C^{\Delta T}}{L_2 R_H^{\Delta T}}\simeq 0.14$, in accordance with the asymptotic shifts of Fig. \ref{fig:DTCNc}. The shifts of the lines in Fig. \ref{fig:SNGTc} are $\frac{G_T}{\delta S_c^{SN}}=(L_3R_C^V)^{-1}\simeq 2.59$, $\frac{T^2G_T}{\delta S_h^{SN}}=(L_4R_H^V-4)^{-1}\simeq -0.41$.

\begin{figure}[t!]
\vspace{-0.75cm}
\center
\hspace*{-0.75cm}
\includegraphics[width=1.25\linewidth]{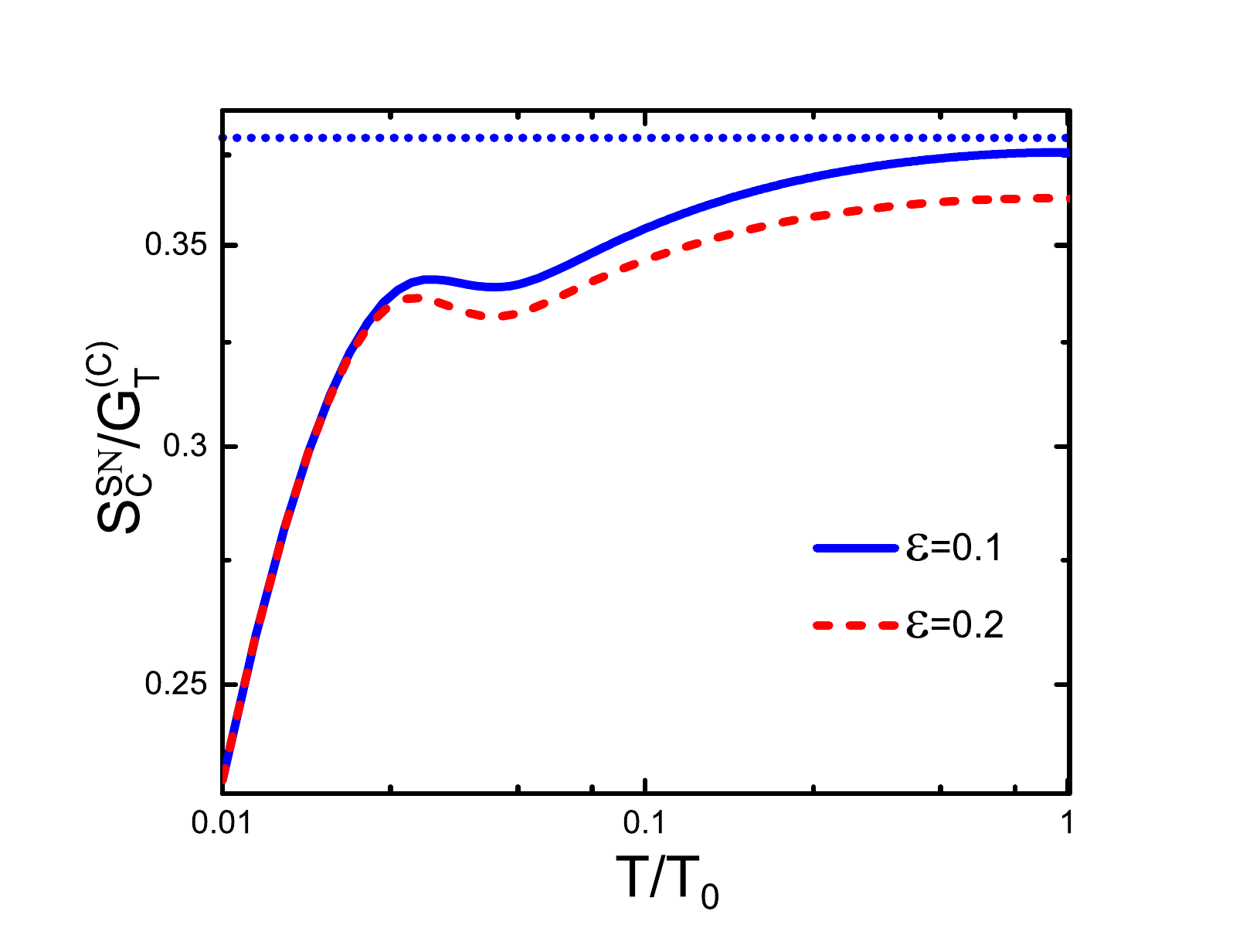} 
\vspace{-1cm}
\caption{\textit{Log-log} plot of the ratio between charge shot noise and thermoelectric coefficient in the conformal limit ($N\rightarrow \infty$) for both elastic and inelastic processes. $\lambda/T_0=0.01$, $E_C/T_0=0.1$. Solid blue line: $\mathcal{E}=0.1$; dashed red line: $\mathcal{E}=0.2$. Blue dotted line corresponds to saturation value  $0.38$ of this ratio for elastic processes for $\mathcal{E}=0.1$ at large temperatures.}
\label{fig:ratio_ANEC}
\end{figure}

In Fig. \ref{fig:ratioSNc}, we demonstrate that, as long as the charging energy is negligible, the ratio of the charge shot noise and the thermoelectric coefficient is temperature-independent both in conformal and Schwarzian regimes. For the illustrative purposes, we make the charging energy negligible comparing to the Schwarzian energy scale, so in both regimes only elastic processes matter. There is a small difference between saturation values for these ratios within the conformal and Schwarzian regimes. It has the following origin: the crossover between the Schwarzian and conformal behavior introduces an additional energy scale to the Green's functions, so the universal single parameter scaling of the Green's functions is, strictly speaking, broken close to $E_{Sch}$. Moving away from this energy scale in either direction, one gets deep into either conformal or Schwarzian regime and these nonuniversality corrections are negligible. Additionally, the crossover between two regimes is relatively smooth comparing to the one steamming from the Coulomb blockade (the former effects into a change of the power-law scaling, while the latter effects into a power-to-exponential transition), so the effects of this universality-breaking scale are small. The explicit form of the two-point Green's function around this crossover is given in \cite{Bagrets2016}. The saturation value in the $\mathcal{E}\rightarrow 0$, $E_C/T\rightarrow 0$ limits in the conformal limit is $\frac{\delta S^{SN}_c}{G_T}=L_3R_C^V\simeq 0.39$, obtained from the general theory. This prediction closely matches the numerical values at finite $\mathcal{E}$ given in the caption of Fig. \ref{fig:ratioSNc}. For the Schwarzian regime, this analytical value is $\simeq 0.338$. The saturation values for all ratios are additionally affected by the spectral asymmetry $\mathcal{E}$. For small $\mathcal{E}$ the corrections are $\sim \mathcal{E}^2$, as shown in Fig. \ref{fig:ratiosSG_all} (since only even powers of $\mathcal{E}$ contribute to symmetric coefficients and only odd powers to antisymmetric ones). At $\mathcal{E}\rightarrow 0$, we have $\frac{\delta S_{c}^{SN}}{G_T}=L_3R_C^V\simeq 0.39$, $\frac{T^2\delta S_c^{SN}}{\delta S_{h}^{SN}}=\frac{L_3R_C^V}{L_4R_H^V-4}\simeq -0.16$, $\frac{T^2\, \delta S_c^{\Delta T}}{\delta S_{h}^{\Delta T}}=\frac{L_1R_C^{\Delta T}}{L_2 R_H^{\Delta T}}\simeq 0.14 $,  $\frac{T^2G}{\delta S_h^{\Delta T}}=\frac{1}{L_2R_H^{\Delta T}}\simeq 0.21$.

Deep into the Coulomb blockade regime, non-elastic processes become dominant for the symmetric transport and noise coefficients. The Green's functions have again a single parametric scaling with temperature, restoring universality of the ratios between the symmetric coefficients. We demonstrate it in Fig. \ref{fig:ratio_NEC} for the ratio between the charge delta-T noise and the charge conductance for the conformal regime ($N\rightarrow \infty$, though the same considerations hold for the Schwarzian regime as well). At $T\ll E_C$, the ratio is temperature independent and stems entirely from the inelastic contribution, its value is $\frac{\delta S_c^{\Delta T}}{G}=\frac{4}{3}$, as expected from Table \ref{tab:ratios}. At $T\gg E_C$, the ratio is dominated by elastic processes and saturates to the conformal elastic value $\frac{\delta S_c^{\Delta T}}{G}=\frac{2}{3}$. The situation is different for the antisymmetric coefficients: the elastic processes always remain important for them, so there are no universal values anywhere below $T\lesssim E_C$ for the ratios between the thermoelectric coefficient, charge shot noise, and heat shot noise. This is illustrated in Fig. \ref{fig:ratio_ANEC}, where the finite-$\mathcal{E}$ ratio between the charge shot noise and the thermoelectric coefficient in the conformal regime saturates solely in the $\frac{E_C}{T}\ll 1$ regime, $\frac{S^{SN}_c}{G_T}\simeq 0.38$.

Overall, the noise coefficients have the following scalings. For the charge shot noise we have
\begin{align}
\label{SNe}    \delta S^{SN}_c\sim\begin{cases}
        \mathcal{E}T^{-\frac{1}{2}}, \,\,\ E_C\ll T \ll J,\\
        \mathcal{E}T^{-1}e^{-\frac{E_C}{2T}}, \,\,\,  E_{Sch}\ll T\ll E_C,\\
        \mathcal{E}T^{-1}e^{-\frac{E_C}{2T}},\,\,\, T\ll E_{Sch}.
    \end{cases}
\end{align}
Comparing to the thermoelectric coefficient, the charge shot noise exhibits the same scaling at high temperatures, as detailed above, while in the low-temperature limit their ratio scales approximately linear with temperature, as is illustrated in Fig. \ref{fig:ratio_ANEC}.

The heat shot noise behaves as
\begin{align}
\label{SNh}    \delta S^{SN}_h\sim\begin{cases}
        \mathcal{E}T^{\frac{3}{2}}, \,\,\ E_C\ll T \ll J,\\
        \mathcal{E}T^{-1}e^{-\frac{E_C}{2T}}, \,\,\,  E_{Sch}\ll T\ll E_C,\\
        \mathcal{E}T^{-1}e^{-\frac{E_C}{2T}},\,\,\, T\ll E_{Sch};
    \end{cases}
\end{align}
the charge delta-T noise is
\begin{align}
\label{TNc}    \delta S^{\Delta T}_c\sim\begin{cases}
        T^{-\frac{1}{2}}, \,\,\ E_C\ll T \ll J,\\
        T, \,\,\,  E_{Sch}\ll T\ll E_C,\\
        T^{\frac{3}{2}},\,\,\, T\ll E_{Sch};
    \end{cases}
\end{align}
and the heat delta-T noise follows
\begin{align}
\label{TNh}    \delta S^{\Delta T}_h\sim\begin{cases}
        T^{\frac{3}{2}}, \,\,\ E_C\ll T \ll J,\\
        T^3, \,\,\,  E_{Sch}\ll T\ll E_C,\\
        T^{\frac{7}{2}},\,\,\, T\ll E_{Sch}.
    \end{cases}
\end{align}

Before proceeding further, we can note that, naively, there is a contradiction for the shot noise and the charge conductance relation at zero temperature. On one hand, $\delta S_c^{SN}\underset{T\rightarrow 0}{=}G$ as defined in Section \ref{sec:TrCoef}. On the other hand, the relations (\ref{Gscaling}) and (\ref{SNe}) mean that these entities have different parities as functions of the spectral asymmetry $\mathcal{E}$ and have different scalings at low temperatures ($T\ll E_C$). This issue is resolved when one notes that within the linear response theory [which is used for Eqs. (\ref{Gscaling}) and (\ref{SNe})], the limits of zero voltage bias and zero temperature are non-commutative, $\lim_{\Delta V\rightarrow 0}\lim_{T\rightarrow 0}\neq \lim_{T\rightarrow 0}\lim_{\Delta V\rightarrow 0}$ (note that it is zero-temperature limit $T\rightarrow 0$ rather than zero temperature difference limit $\Delta T\rightarrow 0$). One can rewrite the Fano factor as
\begin{align}
    F_c^{SN}(T)=\left.\frac{S_c(\Delta V,\Delta T=0)-S_c^{JN}(T)}{G(T)\Delta V}\right\vert_{\Delta V\rightarrow 0},
\end{align}
in this form, it consistently accounts for all regimes and limits. \\
Within the universality regime for the noise coefficients, the Fano factors also mutually obey some universal relations \cite{Pavlov2025, Nguyen2025}. The product of the Fano factors for the charge current becomes a universal constant. Defining $F_c^{\Delta T}=\frac{\delta S_c^{\Delta T}}{G_T}$, we have in the elastic conformal regime
\begin{align}
F_c^{SN}F_c^{\Delta T}=L_1L_3R_C^{V}R_C^{\Delta T}\simeq 0.26.
\end{align}
A similar universal ratio can be introduced for the heat currents. With $F_h^{SN}=\frac{\delta S_h^{SN}}{T^2G_T}$  and  $F_h^{\Delta T}=\frac{\delta S_h^{\Delta T}}{TG_H}$, we have
\begin{align}
F_h^{SN}F_h^{\Delta T}=\frac{L_2R_H^{\Delta T}}{L_0R_L}\left(L_4R_H^V-4\right)\simeq -5.82.
\end{align}

\section{Discussion and Conclusions}
\label{sec:Conc}
In this article, we presented a unified theory of equilibrium transport in a Sachdev-Ye-Kitaev quantum dot coupled to mesoscopic leads through a tunnel junction. This theory treats the transport coefficients and the introduced \textit{noise coefficients} on the same footing and accounts for all types of noise emerging due to voltage or temperature bias. The theory is developed for the linear response regime, though its generalization beyond the linear response is straightforward. We demonstrated that cross-correlators between different currents (i.e., mixed noise) are expressed through charge-charge or heat-heat correlators, so they do not provide new information about the system. Among the noise coefficients only four are independent. In addition to finding explicit scalings of the noise coefficients as functions of temperature $T$ and spectral asymmetry $\mathcal{E}$, we report the universal relations between the transport and noise coefficients. All of the coefficients can be classified as either symmetric or antisymmetric as functions of the spectral asymmetry parameter (this classification, naturally, holds within linear response and breaks in higher orders \cite{Scheibner2005, Karki2017, Arrachea2025}). All coefficients within each group are universally related as long as the $T$-matrix of the system is governed by a single energy scale within a given window of temperatures. This constitutes generalization of the Wiedemann-Franz (WF) law beyond the ratio between heat and charge conductance to ratios between all symmetric and antisymmetric coefficients. This result is different from other generalized WF laws known in literature. For instance, Ref. \cite{Gnezdilov2016} discussed the WF law adapted for Majorana fermion transport, defining it as a ratio between the charge variance (for Majorana fermions, no charge is transmitted on average) and the heat conductance. Refs. \cite{vanDalum2020, Kiselev2020, Karki2020b, Kiselev2023} discussed the generalized WF law, extending the conventional WF to non-Fermi liquids. In particular, for characterization of charge and heat conductance in multichannel Kondo setups. Ref. \cite{Fang2025} introduced a generalization of the WF law relating transport coefficients of currents beyond linear response. We demonstrate that the WF law is just one of the universal relations (laws) mutually connecting four symmetric coefficients and three antisymmetric coefficients. Overall, there are five independent universal relations, others can be trivially expressed through them.

For the particular case of an SYK QD, these results provide explicit theoretical predictions on experimental signatures of quantum dots realizing the SYK physics. Depending on details of measurements in particular setups, shot noise measurements may be preferable to thermoelectric measurements, as it requires introducing a single bias into the system (e.g., voltage bias for conductance and shot noise measurements), rather that tuning between two biases (voltage and temperature) for thermopower measurements \cite{Kleeorin2019}. For instance, in the regime where the Seebeck coefficient of the SYK dot is related to the Bekenstein-Hawking entropy (which holds as long as Coulomb blockade effects are negligible \cite{Pavlov2020}), the shot noise coefficient is directly proportional to the thermoelectric coefficient and, therefore, can be used for the same measurement of the entropy. Using the analytical approach of Ref. \cite{Pavlov2025} and exact numerical calculations, we evaluated the sets of universal constants associated with the SYK quantum dot in various regimes and distinguishing it from the Fermi liquid. These regimes are directly relevant for ongoing efforts in experimental realizations of the SYK physics in mesoscopic graphene flakes.\\
Beyond the SYK dot, our results may be applied to tunneling spectroscopy probes of strongly correlated metals with Planckian dissipation \cite{Wang2024, Gleis2025}.

\textit{Acknowledgments --} We thank Subir Sachdev for a stimulating discussion. A.I.P. appreciates the hospitality of ICTP Trieste, where part of this work was carried out. The work of M.N.K. is conducted within the framework of the Trieste Institute for Theoretical Quantum Technologies (TQT). 

\bibliography{refSYK}

\appendix
\onecolumngrid
\section{Effective tunneling action}
\label{AppA}
The total action of the system averaged over random realizations $J$ and $\lambda$ on the Keldysh contour ($s, \,s^{\prime}=+,\,-$ are indices of the contour) reads as
\begin{align}
\nonumber    S&=\sum_s\int dt \left[s\bar{a}_{q,s}(\textit{i}\partial_t-\varepsilon_q+\mu_L)a_{q,s}+s\bar{c}_{l,s}(\textit{i}\partial_t+\mu_D)c_{l,s}\right]+\int dt dt^{\prime}\sum_{s, s^{\prime}}ss^{\prime}\frac{\textit{i}J^2}{4N^3}\left.\bar{c}_{i,s}\bar{c}_{j,s}c_{k,s}c_{l,s}\right\vert_{t}\left.\bar{c}_{l, s^{\prime}}\bar{c}_{k,s^{\prime}}c_{j, s^{\prime}}c_{i, s^{\prime}}\right\vert_{t^{\prime}}&\\
 \label{actionFull}   &+\int dt dt^{\prime}\sum_{s, s^{\prime}} ss^{\prime}\frac{\textit{i}\lambda^2}{N}e^{\textit{i}\frac{s\chi(t)-s^{\prime}\chi(t^{\prime})}{2}}e^{\textit{i}\frac{s(\varepsilon_q-\mu_L)\xi(t)-s^{\prime}(\varepsilon_{q}-\mu_L)\xi(t^{\prime})}{2}}\left.\bar{a}_{q,s}c_{l,s}\right\vert_t\left.\bar{c}_{l, s^{\prime}}a_{q, s^{\prime}}\right\vert_{t^{\prime}}e^{-\textit{i}\varphi_s(t)}e^{\textit{i}\varphi_{s^{\prime}}(t^{\prime})}+S_C[\varphi]. &
\end{align}
Note that within this formalism we have Grassmann numbers instead of fermionic operators. We also omitted summations over repeated fermionic indices to keep the notations shorter.
Here we applied a Hubbard-Stratonovich transform to eliminate the quartic term in fermionic fields $E_cn^2$ by introducing a new bosonic field $\varphi$:
\begin{align}
    e^{-\textit{i}E_C(\sum_l\bar{c}_{l}c_{l})^2}=\int D\varphi e^{\frac{\textit{i}}{4E_C}\dot{\varphi}^2+\textit{i}\dot{\varphi}\sum_l\bar{c}_{l}c_{l}}
\end{align}
In turn, we removed the second term $\dot{\varphi}\bar{c}_{l}c_{l}$ by means of the gauge transformation $\tilde{c}_{l}\rightarrow c_{l}e^{-\textit{i}\varphi (t)}$ similar to how it was done in \cite{Kamenev1996, Nazarov1999, Efetov2003, Pavlov2020},
\begin{align}
    \bar{c}_{l}\left(\textit{i}\partial_t+\dot{\varphi}\right) c_l\rightarrow \bar{c}_{l}e^{\textit{i}\varphi}\textit{i}\partial_t \left(c_le^{-\textit{i}\varphi} \right).
\end{align}
The remaining part of the action, $S_C[\varphi]$, is the Coulomb action that depends only on the bosonic field $\varphi$.\\
A random nature of tunneling constants $\lambda_{lq}$ in Eq. (\ref{tunneling}) allowed us to make a substantial simplification in Eq. (\ref{actionFull}): the counting field for heat transport is diagonal with respect to $q$, $q^{\prime}$ momenta. It is worth mentioning that it is not so in general. Indeed, due to the nondiagonal contribution, the heat current is nonlocal, and a proper treatment of the heat transport becomes a substantial problem \cite{Fazio1998}. A physical reason for this random structure of tunneling coefficients is that for the irregularly shaped graphene flake wave functions of effective fermionic modes have random rather than deterministic overlaps \cite{Altland2019}.

Now we proceed to integrating out $\bar{c_l}$, $c_l$ fermion fields using the large-$N$ saddle-point solution. We consider only the replica diagonal saddle point, since the off-diagonal terms are strongly suppressed in all regimes \cite{Wang2019, Shackleton2024}. By the usual prescription \cite{Song2017, Can2019, Cheipesh2020} in real time, we introduce $G_{s, s^{\prime}}(t, t^{\prime})=-\frac{\textit{i}}{N}\sum_lc_{l,s}(t)\bar{c}_{l, s^{\prime}}(t^{\prime})$ and use the functional $\delta$-function
\begin{align}
    1=\int D[G]\delta\left(G_{s^{\prime},s}(t^{\prime},t)-\frac{i}{N}\sum_l\bar{c}_{l,s}(t)c_{l, s^{\prime}}(t^{\prime})\right)=\int D[G] D[\Sigma]e^{N\int dt dt^{\prime}\Sigma_{s, s^{\prime}}(t, t^{\prime})\left[G_{s^{\prime},s}(t^{\prime},t)-\frac{i}{N}\sum_i\bar{c}_{i,s}(t)c_{i, s^{\prime}}(t^{\prime})\right]}
\end{align}
to insert it into the action (\ref{actionFull}) above. We obtain
\begin{align}
  \nonumber  S& =S_C[\varphi]+\sum_s\int dt \,s\bar{a}_{q,s}(\textit{i}\partial_t-\varepsilon_q+\mu_L)a_{q,s}+\sum_{s, s^{\prime}}\int dtdt^{\prime}\bigg[\bar{c}_{i,s}\left(s\delta_{s, s^{\prime}}\delta(t-t^{\prime})(\textit{i}\partial_t+\mu_D)-\Sigma_{s, s^{\prime}}(t,t^{\prime})\right)c_{i,s^{\prime}}& \\
    &\left.+\textit{i}ss^{\prime}\frac{NJ^2}{4}G_{s,s^{\prime}}^2(t,t^{\prime})G_{s^{\prime},s}^2(t^{\prime},t)-\textit{i}N\Sigma_{s,s^{\prime}}(t,t^{\prime}) G_{s^{\prime},s}(t^{\prime},t)-\lambda^2 G_{s,s^{\prime}}(t,t^{\prime})\Lambda^{(\chi,\xi)}_{s,s^{\prime}}\bar{a}_{q,s}(t)a_{q,s^{\prime}}(t^{\prime})D^{(\varphi)}_{s,s^{\prime}}(t,t^{\prime})\right],&
\end{align}
where we denoted $\Lambda^{(\chi,\xi)}_{s,s^{\prime}}\equiv ss^{\prime}e^{\textit{i}\frac{s\chi-s^{\prime}\chi}{2}}e^{\textit{i}\frac{\left(s\xi-s^{\prime}\xi\right)(\varepsilon_q-\mu_L)}{2}}$, $D^{(\varphi)}_{s,s^{\prime}}(t,t^{\prime})\equiv e^{-\textit{i}\varphi_s(t)}e^{\textit{i}\varphi_{s^{\prime}}(t^{\prime})}$.

After the integration over the $\bar{c},\, c$ fields, the saddle-point conditions 
\begin{align}
    \frac{\delta S}{\delta G_{s,s^{\prime}}(t,t^{\prime})}=0, \,\,\, \frac{\delta S}{\delta \Sigma_{s,s^{\prime}}(t,t^{\prime})}=0
\end{align}
give us 
\begin{align}
 &\left(\sigma^z(\omega+\mu)-\hat{\Sigma}(\omega)\right)^{-1}_{s, s^{\prime}}=G_{s,s^{\prime}}(\omega),&\\
 &ss^{\prime}J^2G^2_{s,s^{\prime}}(t-t^{\prime})G_{s^{\prime},s}(t^{\prime}-t)=\Sigma_{s,s^{\prime}}(t-t^{\prime})-\textit{i}\frac{\lambda^2}{N}\sum_q \bar{a}_{q,s}(t)a_{q,s^{\prime}}(t^{\prime})\Lambda_{s,s^{\prime}}^{(\chi, \xi)}D^{(\varphi)}_{s,s^{\prime}}(t,t^{\prime}),&
\end{align}
where we used $\sigma^z_{s,s^{\prime}}\equiv s\delta_{s,s^{\prime}}$, and denoted $\hat{\Sigma}=\begin{pmatrix}
    \Sigma_{++} & \Sigma_{+-} \\
    \Sigma_{-+} & \Sigma_{--}
\end{pmatrix}$.

The shift of the saddle point due to the tunneling contact back-action is suppressed as $\frac{1}{N}$ relative to other terms. In the large$-N$ limit, we can neglect this back-action and use the standard SYK saddle-point solution even in the strong tunneling regime. This argument holds both in the conformal and Schwarzian regimes. Note that the chemical shift $\mu$ simply corresponds to the SYK Green's function phase $G(t)e^{\textit{i}\mu t}$, in the frequency representation, it can be accounted by the spectral asymmetry parameter $\mathcal{E}$ \cite{Sachdev2015}. In the Schwarzian regime, the saddle-point Green's function acquires dependence on the fluctuating Goldstone field $h$, $G_{s,s^{\prime}}(t,t^{\prime})\rightarrow G^{(h)}_{s,s^{\prime}}(t,t^{\prime})$, which gives an additional contribution $S_{Sch}[h]$ (soft mode action) to the full action. This soft mode action is additive to the Coulomb action $S_C[\varphi]$ up to $1/N$-small corrections \cite{Altland2019}.
At the saddle point, the action simplifies to
\begin{align}
    S=S_C[\varphi]+S_{Sch}[h]+\sum_{s,s^{\prime}}\int dtdt^{\prime}\sum_q \bar{a}_{q,s}\left(s\delta_{s,s^{\prime}}\delta(t-t^{\prime})(\textit{i}\partial_t-\varepsilon_q)-\lambda^2 \mathcal{G}^{(h,\varphi)}_{s,s^{\prime}}(t,t^{\prime})\Lambda_{s,s^{\prime}}^{(\chi, \xi)}\right)a_{q,s^{\prime}}.
\end{align}
We denoted here $\mathcal{G}^{(h,\varphi)}_{s,s^{\prime}}(t,t^{\prime})\equiv G^{(h)}_{s,s^{\prime}}(t,t^{\prime})D^{(\varphi)}_{s,s^{\prime}}(t,t^{\prime})$. Now, let us integrate the $\bar{a},\, a$ fields. This integral is Gaussian, and can be evaluated exactly. The effective action becomes
\begin{align}
    \textit{i}S=\textit{i}S_c[\phi]+\textit{i}S_{Sch}[h]+\int dt dt^{\prime}\sum_q \Tr \log\left[\sigma^z\delta(t-t^{\prime})(\textit{i}\partial_t-\varepsilon_q)-\lambda^2\mathcal{G}^{(h,\varphi)}_{s,s^{\prime}}(t,t^{\prime})\Lambda_{s,s^{\prime}}^{(\chi,\xi)} \right].
\end{align}
Further simplification gives us
\begin{align}
\label{Sapp}    \textit{i}S=\textit{i}S_C[\varphi]+\textit{i}S_{Sch}[h]+\int \frac{d\varepsilon}{2\pi} \Tr \log \left[1-\lambda^2 \mathcal{G}^{(h,\varphi)}_{s,s^{\prime}}(\varepsilon)\Lambda_{s,s^{\prime}}^{(\chi,\xi)}(\varepsilon)Q_{s^{\prime},s}(\varepsilon)\right],
\end{align}
where $Q_{s^{\prime},s}(\varepsilon)$ is the free-fermion Green's function, $\Lambda_{s,s^{\prime}}^{(\chi,\xi)}(\varepsilon)\equiv ss^{\prime}e^{\textit{i}\frac{s\chi-s^{\prime}\chi}{2}}e^{\textit{i}\frac{\left(s\xi-s^{\prime}\xi\right)(\varepsilon-\mu_L)}{2}}$. Note that the resulting action is local in frequency, despite the Coulomb correlator $D_{s,s^{\prime}}(t,t^{\prime})$ appears to be non-local. This simplification became possible because its expectation value and higher moments depend on the time difference, $t-t^{\prime}$ \cite{Altland2019}. Additionally, in Eq. (\ref{Sapp}), we used $Q_{s^{\prime},s;q}(\varepsilon)=\frac{1}{\rho_F}Q_{s^{\prime},s}(\varepsilon)\delta(\varepsilon-\varepsilon_q)$ [$\rho_F=\sum_q \delta(\varepsilon-\varepsilon_q)$ is the density of states of free fermions] to account for the summations over momenta $q$ and rewrite $\Lambda_{s,s^{\prime}}^{(\chi,\xi)} \rightarrow \Lambda_{s,s^{\prime}}^{(\chi,\xi)}(\varepsilon)$ ($Q_{s^{\prime},s;q}(\varepsilon)$ is the momentum-resolved Green's function) \cite{Karki2018}. 

We now define $\hat{\mathcal{G}}^{(h,\varphi)}(\varepsilon,\chi,\xi)\equiv\begin{pmatrix}
   \mathcal{G}^{(h,\varphi)}_{++}(\varepsilon) & -e^{\textit{i}\chi+\textit{i}\xi(\varepsilon-\mu_L)} \mathcal{G}^{(h,\varphi)}_{+-}(\varepsilon)\\
  -e^{-\textit{i}\chi-\textit{i}\xi(\varepsilon-\mu_L)} \mathcal{G}^{(h,\varphi)}_{-+}(\varepsilon) & \mathcal{G}^{(h,\varphi)}_{--}(\varepsilon)
\end{pmatrix}$, $\hat{Q}(\varepsilon)\equiv\begin{pmatrix}
    Q_{++}(\varepsilon) & Q_{+-}(\varepsilon)\\
    Q_{-+}(\varepsilon) & Q_{--}(\varepsilon)
\end{pmatrix}$. Then the effective action describing tunneling dynamics is given by
\begin{align}
 \label{SeffEx}   \textit{i}S_{eff}[\chi, \xi]=\int \frac{d\varepsilon}{2\pi} \Tr \log \left[1-\lambda^2 \hat{\mathcal{G}}^{(h,\varphi)}(\varepsilon,\chi,\xi)\hat{Q}(\varepsilon)\right].
\end{align}
This is the only term in the full action that depends on the counting fields, so it fully describes the statistics of charge and heat transfer. \\
If one focuses on the conformal limit with negligible charging energy, the expectation values for any moments of Eq. (\ref{SeffEx}) can be calculated exactly (as long as the tunneling coupling $\lambda$ is small enough to disregard the saddle-point shift) \cite{Gnezdilov2018}. This is no longer the case when one accounts for the soft mode fluctuations and charging effects. Then the generating functional is given by 
\begin{align}
    Z[\chi,\xi]=\int D[h]D[\varphi]e^{\textit{i}(S_C[\varphi]+S_{Sch}[h]+S_{eff}[\chi,\xi])}
\end{align}
In this case one has to expand Eq. (\ref{SeffEx}) with respect to small $\lambda^2$, and average it with respect to $h$, and $\varphi$ fields for any chosen momentum (i.e., current or noise) order by order in $\lambda$. This procedure is usually done in the leading order in $\lambda^2$, though Eq. (\ref{SeffEx}) consistently allows this expansion and subsequent calculation up to any order in $\lambda^2$.\\
In the following, we execute this procedure in the leading $\lambda^2$ order, and express the equilibrium observables in terms of the density of states for the SYK dot in all considered regimes. At low temperatures, due to strong Coulomb blockade, inelastic processes coming from the $\lambda^4$ terms become relevant, but they also can be directly incorporated as corrections to the T-matrix of the dot. In the second order, the effective action reads as
\begin{align}
    \textit{i}S_{eff}[\chi, \xi]=-\lambda^2\int \frac{d\varepsilon}{2\pi} \Tr \left[\left\langle\hat{\mathcal{G}}^{(h,\varphi)}(\varepsilon,\chi,\xi)\right\rangle_{h, \varphi}\hat{Q}(\varepsilon)\right].
\end{align}

\section{Current and noise}
\label{AppB}
Here, we explicitly express transport coefficients and noise through the density of states of the dot and the lead.\\
Let us start with rotating Keldysh Green's functions to the triagonal form \cite{Kamenev2011, Song2017}
\begin{align}
    \mathcal{G}^{+-}=\frac{1}{2}\left(\mathcal{G}^A-\mathcal{G}^R+\mathcal{G}^K\right), \,\,\, \mathcal{G}^{-+}=\frac{1}{2}\left(\mathcal{G}^R-\mathcal{G}^A+\mathcal{G}^K\right), 
\end{align}
and the same expressions for $Q$. In Eq. (\ref{cumulants}), the terms in square brackets become
\begin{align}
\mathcal{G}^{+-}Q^{-+}-\mathcal{G}^{-+}Q^{+-}=\frac{1}{2}\left(\mathcal{G}^A-\mathcal{G}^R \right)Q^K-\frac{1}{2}\mathcal{G}^K\left(Q^A-Q^R\right),\\
\mathcal{G}^{+-}Q^{-+}+\mathcal{G}^{-+}Q^{+-}=\frac{1}{2}\mathcal{G}^KQ^K-\frac{1}{2}\left(\mathcal{G}^A-\mathcal{G}^R\right)\left(Q^A-Q^R\right).
\end{align}
Further, $\mathcal{G}^R(\varepsilon)-\mathcal{G}^A(\varepsilon)=-2\pi\textit{i}\rho_{SYK}(\varepsilon)$, $\mathcal{G}^K(\varepsilon)=-2\pi\textit{i}\rho_{SYK}(\omega)\left(1-2n_{SYK}(\varepsilon)\right)$, and the same relations hold for $Q(\varepsilon)$ functions; $\rho_{SYK/F}(\varepsilon)=-\frac{1}{\pi}Im G^R(\varepsilon)$ are the density of states on the SYK dot/Fermi-liquid lead, $n(\varepsilon)=(e^{\frac{\varepsilon-\mu_L}{T}}+1)^{-1}$ is the Fermi-Dirac function. We further use that for the Fermi liquid $\rho_{F}(\varepsilon)=\frac{1}{2\pi v_F}$, where $v_F$ is the Fermi velocity. All kinds of current and noise are explicitly given by 
\begin{align}
\label{currents}    &I_{c/h}=\frac{1}{v_F}\int d\varepsilon \left(\varepsilon-V_L\right)^{n}\rho_{SYK}(\varepsilon)\left[n_{F}(\varepsilon-V_L, T_L)-n_{SYK}(\varepsilon-V_D, T_D)\right],&\\
\label{noises}    &S_{c/m/h}=\frac{1}{v_F}\int d\varepsilon \left(\varepsilon-V_L\right)^l\rho_{SYK}(\varepsilon)\left[n_{F}(\varepsilon-V_L, T_L)+n_{SYK}(\varepsilon-V_D, T_D)-2n_{F}(\varepsilon-V_L, T_L)n_{SYK}(\varepsilon-V_D, T_D)\right],&
\end{align} 
where $n=0,1$ for charge and heat current, correspondingly; $l=0,1,2$ for charge, mixed, and heat noise, correspondingly. $V_{D,L}$ are voltages applied to the dot and the lead, $T_{D,L}$ are temperatures of the dot and the lead (e.g., $\mu_D=V_D$). We also included the $\lambda^2$ coupling constant into the $\rho_{SYK}(\varepsilon)$.

The electric conductance (for $V_D=0$, $V_L=\Delta V$) is
\begin{align}
   G= \left.\frac{\partial I_c}{\partial\Delta V}\right\vert_{\Delta T=0}=\frac{1}{4T v_F}\int d\varepsilon \rho_{SYK}(\varepsilon)\frac{1}{\cosh^2\frac{\varepsilon}{2T}}.
\end{align}
The thermoelectric coefficient $G_T$ (for $V_D=V_L=0$, $T_D=T$, $T_L=T+\Delta T$) is
\begin{align}
\label{GTexpr}   G_T=\left.\frac{\partial I_c}{\partial\Delta T}\right\vert_{\Delta V= 0}=\frac{1}{4 T^2 v_F}\int d\varepsilon \rho_{SYK}(\varepsilon) \frac{\varepsilon}{\cosh^2\frac{\varepsilon}{2T}}.
\end{align}
For the heat current, one has $G_H$ (for $V_D=V_L=0$, $T_D=T$, $T_L=T+\Delta T$), related to the thermal conductance,
\begin{align}
    G_H=\left.\frac{\partial I_h}{\partial\Delta T}\right\vert_{\Delta V= 0}=\frac{1}{4T^2 v_F}\int d\varepsilon \rho_{SYK}(\varepsilon)\frac{\varepsilon^2}{\cosh^2\frac{\varepsilon}{2T}}.
\end{align}
The equilibrium (Johnson-Nyquist) thermal noise (at $V_D=V_L=0$, $T_D=T_L=T$) is 
\begin{align}
\label{JNnoise}    S^{JN}_{l}(T)=\frac{1}{2v_F}\int d\varepsilon \rho_{SYK}(\varepsilon) \frac{\varepsilon^l}{\cosh^2\frac{\varepsilon}{2T}}.
\end{align}
Here and in the following, the subscript $l$ stands for notations $c,m,h$, and the power $l=0,1,2$, correspondingly.\\ 
For delta-T noise, we consider $V_D=V_L=0$, $T_D=T$, $T_L=T+\Delta T$ and assume deviations from equilibrium to be small, so only linear terms in $\Delta T$ matter. Delta-T noise $\delta S_l^{\Delta T}$ becomes
\begin{align}
\label{dTnoise}   \delta S_{l}^{\Delta T}=\left.\frac{\partial S_{l}(\Delta V, \Delta T)}{\partial \Delta T}\right\vert_{\Delta V=0}
    =\frac{1}{4T^2 v_F}\int d\varepsilon \rho_{SYK}(\varepsilon)\frac{\varepsilon^{1+l} \sinh\frac{\varepsilon}{2T}}{\cosh^3\frac{\varepsilon}{2T}}.
\end{align}
To find the shot noise $\delta S_l^{SN}$, we set $V_D=0$, $V_L=\Delta V$, $T_L=T_R=T$,
\begin{align}
\label{SNoise}   \delta S_l^{SN}=\left.\frac{\partial S_l(\Delta V, \Delta T)}{\partial \Delta V}\right\vert_{\Delta T=0}=\frac{1}{4 T v_F}\int d\varepsilon \rho_{SYK}(\varepsilon) \frac{\varepsilon^l\sinh\frac{\varepsilon}{2T}}{\cosh^3\frac{\varepsilon}{2T}} -\frac{l}{2v_k}\int d\varepsilon \rho_{SYK}(\varepsilon)\frac{\varepsilon^{|l-1|}}{\cosh^2\frac{\varepsilon}{2T}} .
\end{align}
One can rewrite the expressions above in a more compact form by introducing transport integrals $\mathcal{L}_n$ \cite{Costi2010, Karki2020} and generalizing them to noise integrals $\mathcal{N}_n$, defined as
\begin{align}
    &\mathcal{L}_n=\frac{1}{4T v_F}\int_{-\infty}^{\infty}d\varepsilon \rho_{SYK}(\varepsilon)\frac{\varepsilon^n}{\cosh^2\left(\frac{\varepsilon}{2T}\right)}, \, n=0,1,2;&\\
    &\mathcal{N}_n=\frac{1}{4T v_F}\int_{-\infty}^{\infty}d\varepsilon \rho_{SYK}(\varepsilon)\frac{\varepsilon^n\sinh\left(\frac{\varepsilon}{2T}\right)}{\cosh^3\left(\frac{\varepsilon}{2T}\right)}, \, n=0,1,2,3.&
\end{align}
We have then 
\begin{align}
    &G=\mathcal{L}_0,\,\,\, G_T=\frac{1}{T}\mathcal{L}_1,\,\,\, G_H=\frac{1}{T}\mathcal{L}_2;&\\
    &S_c^{JN}=2T\mathcal{L}_0,\,\,\,\delta S_c^{SN}=\mathcal{N}_0,\,\,\, \delta S_c^{\Delta T}=\frac{1}{T}\mathcal{N}_1;&\\
    &S_m^{JN}=2T\mathcal{L}_1,\,\,\,\delta S_m^{SN}=\mathcal{N}_1-S_c^{JN},\,\,\, \delta S_m^{\Delta T}=\frac{1}{T}\mathcal{N}_2;&\\
    &S_h^{JN}=2T\mathcal{L}_2,\,\,\,\delta S_h^{SN}=\mathcal{N}_2-2S_m^{JN},\,\,\, \delta S_h^{\Delta T}=\frac{1}{T}\mathcal{N}_3.&\\    
    \end{align} 
In general, the density of states of a dot in the expressions above can be expressed through the imaginary part of the T-matrix in Matsubara representation $\mathcal{T}$, as is derived in \cite{Matveev2002},
\begin{align}
    \rho(\varepsilon)=-\frac{1}{\pi}\cosh\frac{\varepsilon}{2T}\int_{-\infty}^{\infty}dt \mathcal{T}\left(\frac{1}{2T}+\textit{i}t\right)e^{\textit{i} \varepsilon t}.
\end{align}
We put this expression into the formulas above and change the order of integration, obtaining for transport coefficients
\begin{align}
   &G=-\frac{1}{2v_F}\int_{-\infty}^{\infty} dt \frac{1}{\cosh(\pi T t)}\mathcal{T}\left(\frac{1}{2T}+\textit{i}t\right),&\\
    &G_T=-\frac{\textit{i}\pi}{2v_F}\int_{-\infty}^{\infty} dt \frac{\sinh (\pi T t)}{\cosh^2 (\pi T t)}\mathcal{T}\left(\frac{1}{2T}+\textit{i}t\right),&\\
   &G_H=-\frac{\pi^2 T}{v_F}\int_{-\infty}^{\infty} dt \frac{1}{\cosh^3 (\pi T t)}\mathcal{T}\left(\frac{1}{2T}+\textit{i}t\right)-T\pi^2 G.&
\end{align}

For the charge noise, the same procedure gives us
\begin{align}
    &S^{JN}_c=-\frac{T}{v_F}\int_{-\infty}^{\infty}dt\frac{1}{\cosh(\pi T t)}\mathcal{T}\left(\frac{1}{2T}+\textit{i}t\right)=2TG,&\\
    &\delta S_c^{SN}=-\frac{\textit{i}T}{v_F}\int_{-\infty}^{\infty}dt\frac{t}{\cosh(\pi T t)}\mathcal{T}\left(\frac{1}{2T}+\textit{i}t\right),&\\
    &\delta S_c^{\Delta T}=\frac{\pi T}{v_F}\int_{-\infty}^{\infty}dt \frac{\, t\sinh\left(\pi T t\right)}{\cosh^2\left(\pi T t\right)} \mathcal{T}\left(\frac{1}{2T}+\textit{i}t\right)+2G.&
\end{align}
For the heat noise, we have
\begin{align}
    &S_h^{JN}=-\frac{\pi^2 T^3}{v_F}\int_{-\infty}^{\infty}dt\left(\frac{2-\cosh^2\left(\pi T t\right)}{\cosh^3\left(\pi T t\right)}\right) \mathcal{T}\left(\frac{1}{2T}+\textit{i}t\right)=2T^2G_H,\\
    &\delta S_h^{SN}=-\frac{2\textit{i}\pi^2 T^3}{v_F}\int_{-\infty}^{\infty}dt \frac{t}{\cosh^3\left(\pi T t\right)}\mathcal{T}\left(\frac{1}{2T}+\textit{i}t\right)-\pi^2T^2\,\delta S_c^{SN},&\\
    &\delta S_h^{\Delta T}=
    \frac{\pi^3T^3}{v_F}\int_{-\infty}^{\infty}dt \frac{\,t \sinh\left(\pi Tt\right)\left(5-\sinh^2\left(\pi Tt\right)\right)}{\cosh^4\left(\pi Tt\right)}\mathcal{T}\left(\frac{1}{2T}+\textit{i}t\right)+6TG_H.&
 \end{align}   
 The mixed (cross-correlation) noise is not independent from the quantities found above. A straightforward comparison of Eq. (\ref{JNnoise}) with Eq. (\ref{GTexpr}), and Eq. (\ref{dTnoise}) with Eq. (\ref{SNoise}) gives 
\begin{align}
    &S_m^{JN}=-\frac{\textit{i}\pi T^2}{v_F}\int_{-\infty}^{\infty} dt \frac{\sinh\left(\pi Tt\right)}{\cosh^2\left(\pi Tt\right)}\mathcal{T}\left(\frac{1}{2T}+\textit{i}t\right)=2T^2G_T,&\\
    &\delta S_m^{SN}=T\,\delta S_c^{\Delta T}-S_c^{JN},&\\      
    &\delta S_h^{SN}=T\delta S_m^{\Delta T}-2S_m^{JN}.&
\end{align} 
Now, let us analyze the Fano factor for the charge noise and charge current in the limit of zero temperature. In this case, Fermi-Dirac functions in Eqs. (\ref{currents}) and (\ref{noises}) turn into the Heaviside functions $n_{F/SYK}(\varepsilon-V_{D/L})\longrightarrow \theta (\varepsilon-V_{D/L}-\mu)$. The Heaviside function is idempotent, meaning $\theta^2(x)=\theta(x)$, therefore, the ratio of expressions for the charge noise and charge current becomes unity: 
\begin{align}
F^{SN}\equiv\frac{S_c}{I_c}\underset{T\rightarrow 0}{=} 1.
\end{align}
\end{document}